\documentclass[%
 aip, pof,
amsmath,amssymb,
preprint,
]{revtex4-1}

\usepackage{graphicx}
\usepackage[dvipsnames]{xcolor} 
\usepackage{dcolumn}
\usepackage{bm}

\usepackage[utf8]{inputenc}
\usepackage[T1]{fontenc}
\usepackage{mathptmx}
\usepackage{etoolbox}

\usepackage{derivative}
\usepackage{subfigure}
\usepackage{svg}
\usepackage{cleveref}
\usepackage{mathtools}
\usepackage{cleveref}

\makeatletter
\def\@email#1#2{%
 \endgroup
 \patchcmd{\titleblock@produce}
  {\frontmatter@RRAPformat}
  {\frontmatter@RRAPformat{\produce@RRAP{*#1\href{mailto:#2}{#2}}}\frontmatter@RRAPformat}
  {}{}
}%
\makeatother

\begin{document}


\title{\large{Active Flow Control for Bluff Body under High Reynolds Number Turbulent Flow Conditions Using Deep Reinforcement Learning}}

\author{Jingbo Chen}
\affiliation{Politecnico di Milano, Piazza Leonardo da Vinci 32, 20133 Milano, Italy}
\email{jingbo.chen@mail.polimi.it}


\author{Enrico Ballini}
\affiliation{MOX, Department of Mathematics, Politecnico di Milano, Piazza Leonardo da Vinci 32, 20133 Milano, Italy}

\author{Stefano Micheletti}
\affiliation{MOX, Department of Mathematics, Politecnico di Milano, Piazza Leonardo da Vinci 32, 20133 Milano, Italy}



\begin{abstract}
This study employs Deep Reinforcement Learning (DRL) for active flow control in a turbulent flow field of high Reynolds numbers at $Re=274000$. That is, an agent is trained to obtain a control strategy that can reduce the drag of a cylinder while also minimizing the oscillations of the lift. Probes are placed only around the surface of the cylinder, and a Proximal Policy Optimization (PPO) agent controls nine zero-net mass flux jets on the downstream side of the cylinder. The trained PPO agent effectively reduces drag by $29\%$ and decreases lift oscillations by $18\%$ of amplitude, with the control effect demonstrating good repeatability. Control tests of this agent within the Reynolds number range of $Re=260000$ to $288000$ show the agent's control strategy possesses a certain degree of robustness, with very similar drag reduction effects under different Reynolds numbers. Analysis using power spectral energy reveals that the agent learns specific flow frequencies in the flow field and effectively suppresses low-frequency, large-scale structures. Graphically visualizing the policy, combined with pressure, vorticity, and turbulent kinetic energy contours, reveals the mechanism by which jets achieve drag reduction by influencing reattachment vortices. This study successfully implements robust active flow control in realistically significant high Reynolds number turbulent flows, minimizing time costs (using two-dimensional geometrical models and turbulence models) and maximally considering the feasibility of future experimental implementation. 

\end{abstract}

\maketitle

\section{\label{sec:level1}Introduction}

In aircraft, navigation, and related industries, we often encounter the technical task of minimizing drag. To address these challenges, we usually depend on passive control or active control \cite{GadelHak2000}. An approach to reducing drag in specific cases is by modifying the aerodynamic shape of an object\cite{Viquerat2021110080}, for example, the dimpled design of golf balls \cite{Chowdhury201687}.

For items such as large trucks, trains, wind turbines and aircrafts that have geometric shapes set and where reducing drag leads to a decrease in energy consumption \cite{Aubrun2017, Pereira2017, Zhang2024, Bohn2019}, active control is required after aerodynamic optimization has already been completed. 

Active control can be categorized as open-loop active control \cite{Shahrabi2019} and closed-loop active control. Open-loop active control is easier to set up, but is very unstable when dealing with external disruptions or internal parameter variations. Closed-loop active control is frequently utilized in intricate situations for flow control \cite{Choi2008, ChuanEun2018, Aamo2004, Gautier2015, Leclercq2019}, particularly in the context of Active Flow Control (AFC) \cite{ScottCollis2004237}. 

In AFC, the input consists of physical measurements collected by probes within the flow field, while the output involves altering the flow field using actuators. Establishing a reliable transfer function between input and output is difficult, particularly for turbulent flow fields, and the stability of the transfer function is weak due to several disruptions in real-world settings \cite{Brunton2015}. Therefore, it is crucial to integrate machine learning (ML) for active closed-loop control of the flow field \cite{Mao2022, Lee1997, Ren_Rabault_Tang2021, Ren_Wang_Tang2021, Rabault2019, Fan2020, Li_Zhang_2022, Tang2020, Zheng2021, Brunton2015, Duriez2017}. The integration of machine learning with fluid dynamics is closely related and has been applied in research areas such as flow field reconstruction and prediction \cite{Sekar2019}, turbulence modeling \cite{Duraisamy2019, Thuerey2020}, and flow control \cite{Ren_Wang_Tang_2019, Raibaudo2020, Siddhartha2018, Zhu2019}. Yunfei Li \cite{Li202214} and Steven L. Brunton \cite{Brunton2020} offer insightful overviews of recent advances in the field. Reinforcement Learning (RL) \cite{Sutton2018, Colabrese2017} or Deep Reinforcement Learning (DRL) \cite{Arulkumaran2017} approaches are frequently selected to achieve control. DRL utilizes Neural Networks (NN) to address complicated, high-dimensional problems within the subject of ML, whereas RL utilizes tabular approaches and Q-Learning to develop strategies. Reinforcement learning is not feasible for complex control issues that involve continuous high-dimensional action spaces. Hence, we opt to utilize DRL with Deep Neural Networks (DNN) to address intricate control challenges \cite{Schmidhuber201585, Duan2016, Bohn2019, Bucci2019, Mnih2015, Rabault2020, Rabault_Kuhnle2019}.

Deep Neural Networks in DRL acts as transfer functions in control theory, serving as a black-box model to accurately represent the relationships between inputs and outputs. DNN is well-suited for modeling complicated nonlinear systems, like those needing turbulence control, which are difficult to model using traditional methods. Therefore, substantial computational effort is required for proper fitting of the nonlinear function between inputs and outputs, along with a suitable update approach to update the weights of DNN. Recent research focuses on machine learning-based active flow control at low Reynolds numbers \cite{Ren_Rabault_Tang2021, Rabault2019, Tang2020, Zheng2021}. Researchers such as Feng Ren \cite{Ren_Rabault_Tang2021} and Hongwei Tang \cite{Tang2020} have successfully reduced the drag on bluff bodies under weakly turbulent laminar and turbulent conditions, achieving effective control across various Reynolds numbers. Active flow control at high Reynolds numbers is important in various engineering applications, including reducing drag for cars \cite{Seifert2009, Heinemann2014, Krajnovic2011192}, high-speed trains and wings \cite{Kasmaiee2023}, as well as attitude control for aircraft without rudders. Studies such as those of Zhijie Zhao have utilized synthetic jets for attitude control of such aircraft \cite{Zhao2022} and Enrico Ballini successfully implemented flow control over bluff bodies at higher Reynolds numbers using the PPO algorithm \cite{ballini2023reducingdragbluffbody}.

Optimizing control for high Reynolds number flow configurations using DRL encounters numerous hurdles. Fluid mechanics simulations are time-intensive, and training the network requires detailed flow field data like velocity and pressure, resulting in a huge number of network inputs. The significant time cost for training the agent is related to the high non-linearity between inputs and outputs. When computational resources are limited, it is practical to begin by utilizing simplified models to investigate efficient flow field control. This can involve employing two-dimensional geometric models or a range of turbulence models during the early stages of study. Efficiently exploiting samples is key in DNN's update process. This is because a large amount of samples requires significant data processing, and the majority of deep-reinforcement learning agents use the Gradient Descent Method (GDM) for updates. Increasing the number of samples improves the accuracy of gradient computation, but also leads to a substantial increase in computational workload, requiring a trade-off. It is crucial to reduce the total time cost of agent training in turbulent flow circumstances by examining the agent's ability to learn flow field properties and optimizing control based on features. The stability of the training results can be assessed by examining the reproducibility of the training and the robustness and reaction speed of the selected policy control mechanism after training, focusing on control aspects.

Previous research on flow control with DRL has mainly used the PPO algorithm \cite{Ren_Rabault_Tang2021, Rabault2019, Tang2020, Schulman2017}. This label-free learning technique streamlines data processing and is commonly employed in flow control for many reasons. Flow control is a challenge due to the continuous action space, and the PPO algorithm excels at addressing such complexities, playing a vital role in improving decision-making. Furthermore, PPO includes a clipping mechanism that inhibits sudden changes in actions. PPO has an exploratory tendency that helps prevent getting stuck in local optima, although this can lead to slower convergence rates. Defining the reward function enables the optimization of many objectives at the same time.

In view of these PPO algorithmic features, this article also employs PPO as the basis for its investigation of flow control-based drag reduction. Virtual probes are positioned only on the cylinder surface to prevent interference from upstream probes in the wake from impacting downstream probes, considering practicality in real experimental settings and engineering uses, as well as the significance of experimental-simulation comparisons. Given these limitations, the maximum amount of flow field data is collected, and the flow field is managed using jetting. Xinhui Dong et al. placed sensors around the cylinder to monitor pressure and trained an agent for suction control, experimentally achieving a reduction in the adverse effects of vortex shedding at the rear of the cylinder and suppressing lift fluctuations \cite{Dong2023}. 

The policy developed in this study significantly reduced the drag on the cylinder by 29\% in a flow field at $Re=274000$ and exhibited a very high response speed. This control strategy is also robust and capable of handling variations in Reynolds numbers within a certain range without excessively increasing the oscillation of vortex shedding. The paper is framed as follows. Section~\ref{sec:methodology} introduces the numerical methods and the PPO algorithm, as well as the methodology for coupling the agent with the simulation environment. Section~\ref{sec: result_and_discuss} discusses the trained agent based on the repeatability of the training results, analyzing the impact of control on the flow field through velocity, pressure, and vorticity contour plots. The effectiveness of the control is further examined by comparing and analyzing data and graphs. Section~\ref{sec:conclusion} summarizes the work presented in this paper and describes directions for future research.

\section{Methodology}\label{sec:methodology}
\subsection{Flow Configuration}
\begin{figure*}
\subfigure[]{\includegraphics[width = .5\linewidth]{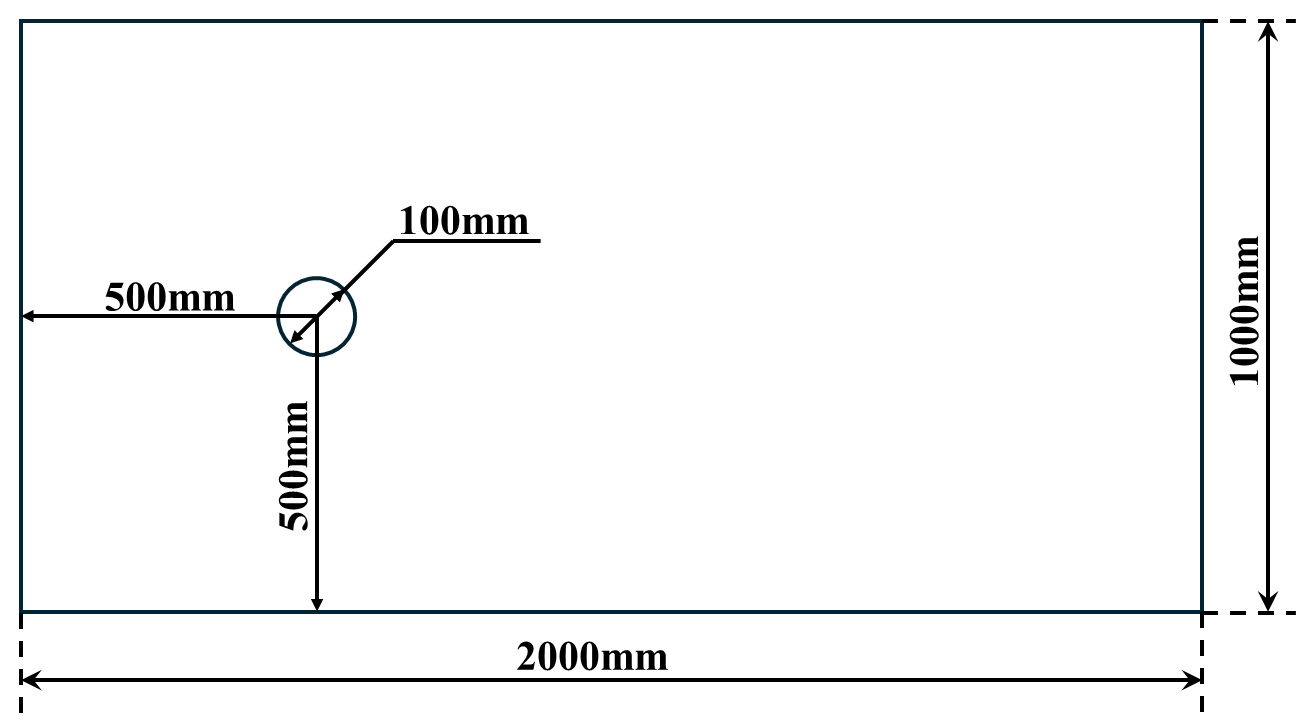}}\hfill
\subfigure[]{\includegraphics[width = .47\linewidth]{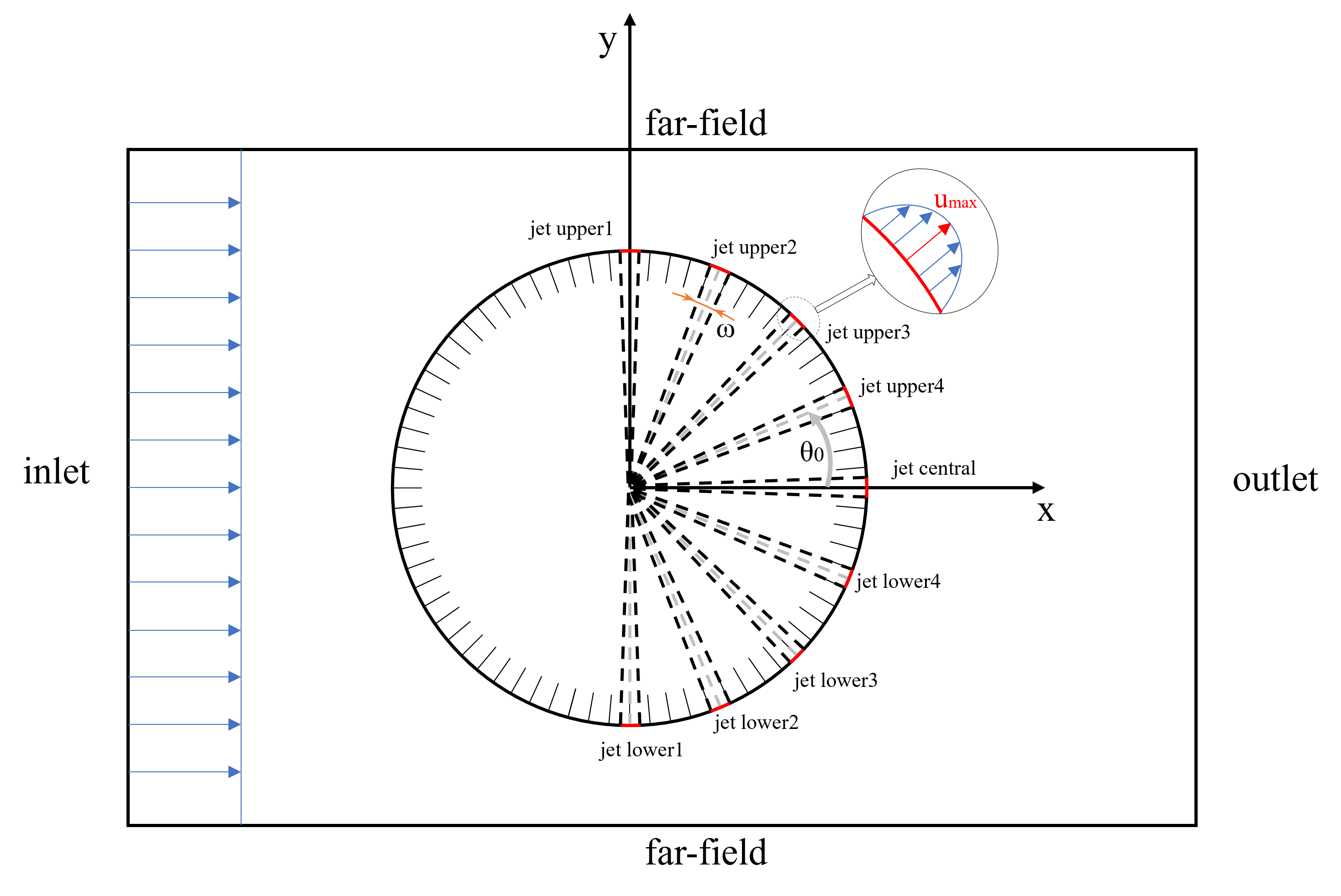}}
\caption{Geometrical configuration of the circular cylinder and the computational domain: (a) The relative position of the cylinder in the computational domain. (b) Schematic of the jets on the cylinder with geometrical dimensions and boundary conditions with the profile of velocity inlet and the profile of jets.}
\label{fig:1}
\end{figure*}

The configuration in this investigation is a turbulence flow past a circular cylinder in a two-dimensional setting with a Reynolds number ($Re$) of $274000$ where $Re=\bar{U}D/\nu$. $\bar{U}$ is the mean velocity of inflow from infinity, $D$ is the diameter of the cylinder, and $\nu$ is the kinematic viscosity of the fluid. 
The diameter of the cylinder is $100mm$ and the domain of the numerical simulation is $2000mm(length) \times 1000mm(height)$, the center of the cylinder is placed $500mm$ away from the inlet boundary and $500mm$ away from the far field boundary (see Fig.~\ref{fig:1}(a)).

For flow control, we use jets to influence the flow field around a cylinder to achieve the control objective. Figure~\ref{fig:1}(b) shows the configuration of nine control jets which are installed symmetrically on the surface of the cylinder with different angular positions relative to the $x$-axis. We can define the position of the jets through $\theta$, $\theta_i (jet upper) = 90^{\circ},67.5^{\circ},45^{\circ},22.5^{\circ}$ with $i=1,2,3,4$ and $\theta_i (jet lower) =-90^{\circ},-67.5^{\circ},-45^{\circ},-22.5^{\circ}$ with $i=1,2,3,4$ and finally the central jet at $0^{\circ}$. The dimension of the jet outlet is limited by the angle $\omega$ and the angle $\omega$ is always equal to $5^{\circ}$. To avoid the discontinuity of velocity when we define the boundary condition of the jets and no-slip wall on the surfaces of the cylinder, we choose the velocity profile at each jet outlet following the expression defined below \cite{Tang2020}:
\begin{equation}
{u=u}_{\max}\cos{\bigg(\frac{\pi}{\omega}(\theta-\theta_0)\bigg)},
\label{eq:1}
\end{equation}
where $u_{\max}$ is limited between $-2m/s$ and $2m/s$ and because it is an incompressible flow, the density of the fluid is constant, which means that the Artificial Neural Network (ANN) delivers the velocities, so the agent controls the velocity of nine jets directly: when the jet velocity is positive, it means blowing; otherwise it means suction. The nine jets are zero net mass flux jets, implying that at any given moment the total flux through the boundaries of these nine jets is zero, eliminating the need for additional mass injection into the flow field.

\subsection{Problem description}

In the present study, the flow is considered viscous and incompressible. The stationary no-slip wall is applied to the cylinder. The governing equations are the two-dimensional Reynolds-averaged Navier-Stokes equations (RANS).


The continuity equation for incompressible flow reads:
\begin{equation}
    \frac{\partial \overline{u_i}}{\partial x_i} = 0.
\end{equation}

The momentum equation for incompressible flow reads:
\begin{equation}
\rho \left( \frac{\partial \overline{u_i}}{\partial t} + \overline{u_j} \frac{\partial \overline{u_i}}{\partial x_j} \right) = -\frac{\partial \overline{p}}{\partial x_i} + \mu \frac{\partial^2 \overline{u_i}}{\partial x_j^2} - \rho\frac{\partial \overline{u_i' u_j'}}{\partial x_j}.
\end{equation}
To close the system, we need to model the Reynolds stresses $\overline{u_i' u_j'}$. In this regard, we adopt the $k-\omega$ SST turbolence model, which can calculate flow separation relatively accurately, according to the research of Wen and Qiu \cite{Wen2017, Langtry2009}. The use of the Large Eddy Simulation (LES) method, which directly resolves large-scale eddies, imposes severe requirements on the mesh and demands more computational resources. The 2D and 3D RANS models still exhibit shortcomings in capturing the drag crisis phenomenon \cite{Singh2005}, but the calculated drag coefficients remain within a reasonable range \cite{Wen2017}. Given the already substantial expense of carrying out numerical simulations in three dimensions, combined with the extensive time needed for training the agent, the task becomes exceedingly challenging. Therefore, we chose to simplify this model to a two-dimensional problem and apply the RANS model. In the present work, we use the SIMPLE scheme to solve the incompressible RANS model problem. The numerical simulation is based on the commercial software ANSYS-Fluent.
Drag, $F_{D}$, and lift, $F_{L}$, can be calculated by integrating the viscous stress and pressure acting on the surface of the cylinder, and in this case, the reaction force contributed by jet propulsion should be considered. The drag force and lift force, the drag coefficient, $C_{D}$, and the lift coefficient, $C_{L}$, are defined below.
\begin{equation}
F_D=\int (\sigma\cdot n_c)\cdot e_x \,dS,
\label{eq:4}
\end{equation}
\begin{equation}
F_L=\int (\sigma\cdot n_c)\cdot e_y \,dS,
\label{eq:5}
\end{equation}
\begin{equation}
C_D=\frac{2F_D}{\rho{\bar{U}}^2D},
\label{eq:6}
\end{equation}
\begin{equation}
C_L=\frac{2F_L}{\rho{\bar{U}}^2D}.
\label{eq:7}
\end{equation}
The stress tensor is denoted by $\sigma$, $n_c$ represents the unit outward normal vector to the outer surface of the cylinder, and $e_x=(1,0)$ and $e_y=(0,1)$.
The Strouhal number, $St$, is a dimensionless number used to represent the frequency of the dominant oscillation of the flow, and is defined as follows:
\begin{equation}
St = \frac{f_s\, D}{\bar{U}}.
\label{eq:8}
\end{equation}
The quantity $f_s$ is a natural frequency, representing the vortex shedding frequency that can be calculated from the periodic oscillation of the lift coefficient, $C_L$.

\subsection{Mesh independence}
\begin{figure}[h]
  \centering
  \includegraphics[width=.8\linewidth]{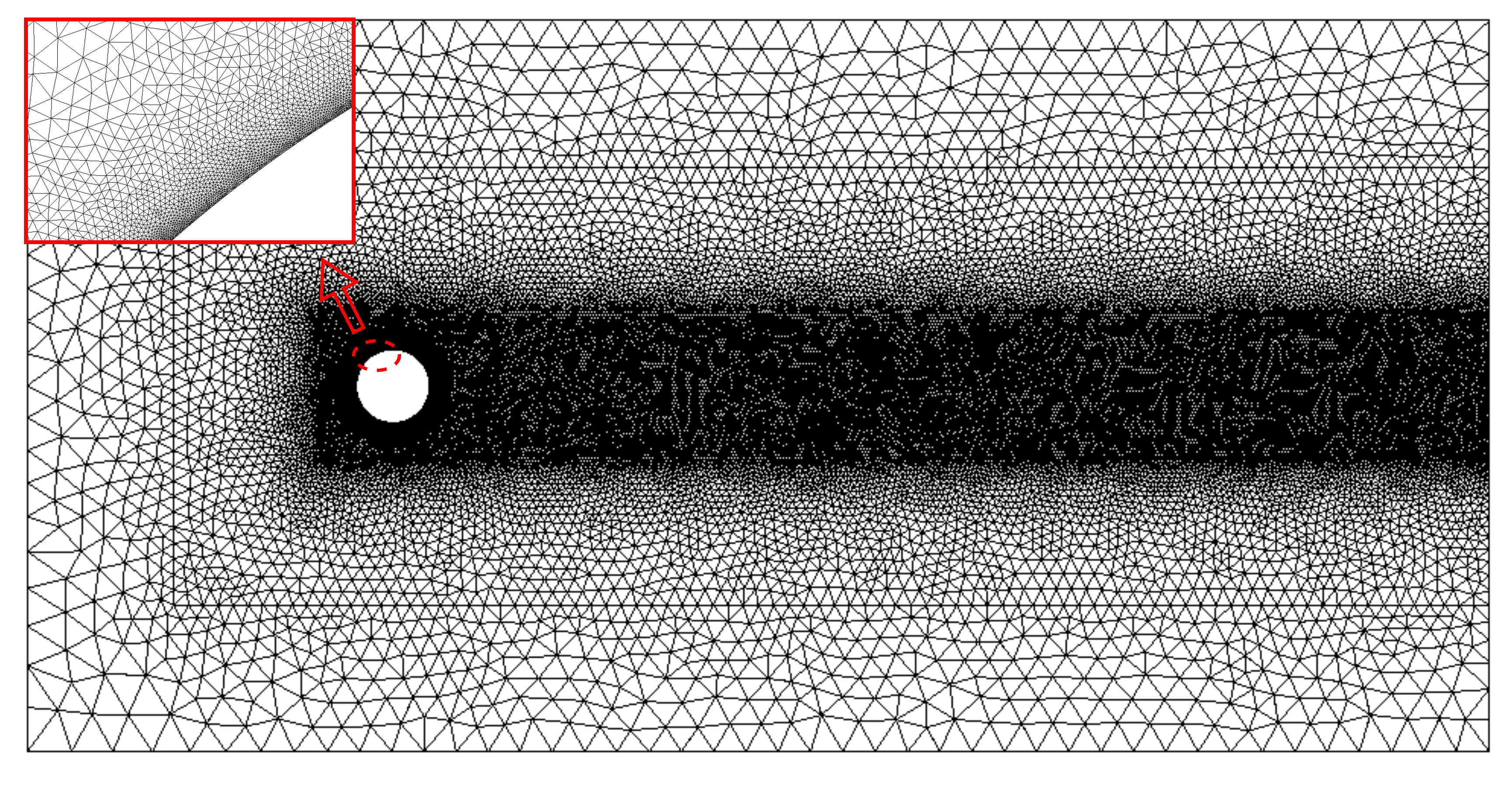}
  \caption{Computational mesh: the innermost mesh is refined and the mesh around the cylinder is highly refined considering the accuracy of the simulation in the boundary layer.}
  \label{fig:2}
\end{figure}

The unstructured mesh used in the simulation is made up of 182612 triangular grids (Fig.~\ref{fig:2}). To further reduce the computational load without compromising the accuracy of the numerical simulations, we divide the computational domain into three nested regions with increasing grid density. In addition, the grid around the cylindrical wall is much more refined because of the boundary layer.
Table~\ref{tab:table1} illustrates that convergence is reasonably assumed with respect to the mesh size and time step. The mean values of $C_D$ for meshes of 182612 and 283061 elements at a time step of $\delta_t = 0.00025s$ vary by just $0.5\%$. Therefore, considering computational efficiency, we opted for the mesh with 182612 elements for this study. 
\begin{table}[h]
\caption{\label{tab:table1} Mesh convergence with mesh resolution and time step size referring drag coefficient for the flow around the circular cylinder at $Re=274000$.}
\begin{ruledtabular}
\begin{tabular}{l|cccc}
$N. elements$ & 28874 & 85849 & 182612 & 283061\\ \hline
$\delta_t=0.001$ & & & & \\ 
$C_D \_ mean$ & 0.52522 & 0.78816 & 0.79094 & \\
$\delta_t=0.0005$ & & & & \\ 
$C_D \_ mean$ & & 0.87778 & 0.84921 & \\
$\delta_t=0.00025$ & & & & \\ 
$C_D \_ mean$ & & 0.84996 & 0.89337 & 0.89754 \\
\end{tabular}
\end{ruledtabular}
\end{table}

\subsection{Deep Reinforcement Learning}
\begin{figure}[h]
    \centering
    \includegraphics[width=0.7\linewidth]{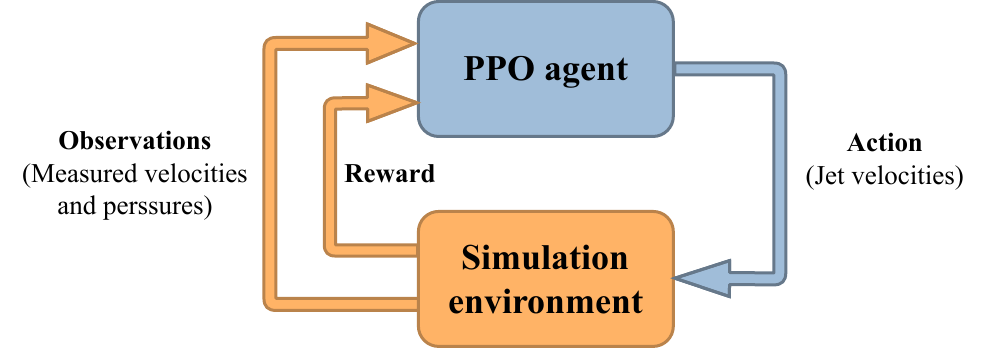}
    \caption{Schematic of the PPO agent closed-loop control.}
    \label{fig:5}
\end{figure}
The basic framework of PPO is illustrated in Fig.~\ref{fig:5}, and the details of the PPO algorithm are presented in Appendix~\ref{sec:appendixA}, while the employed hyper parameters are listed in Appendix~\ref{sec:appendixB}. It is essentially a closed-loop control system where observations and rewards are input variables, and actions are output variables during neural network training. For a pre-trained neural network, inputting an observation allows the agent to quickly select the corresponding policy, which is the action. The training process can be viewed as constructing a transfer function or, more precisely, approximating a transfer function. In general, the agent interacts with the environment, which in our study is a virtual environment corresponding to the simulation of a flow field. The data generated from the flow field simulation is transmitted across a channel to the agent as an observation input. This corresponds to the velocity and pressure information obtained from the probes. Additionally, the lift coefficient and drag coefficient are calculated by simulation to compute the reward input for the PPO agent:
\begin{equation}
  Reward\ =-\left|C_D\right|-0.2\left|C_L\right|. 
\label{eq:31}
\end{equation}
A specific weighting for drag and lift is used to minimize drag without significantly boosting lift, emphasizing the importance of this optimization target in relation to the research objectives. The agent uses observations and rewards to train the DNN and computes actions. This study examines the maximum values of nine jet vectors. The values are used as input for the velocity profile function of the jets, and the equations act as boundary conditions for the numerical simulations. Executing the action in the environment completes a full interaction cycle between the PPO agent and the environment.
\begin{figure}[h]
    \centering
    \includegraphics[width=0.7\linewidth]{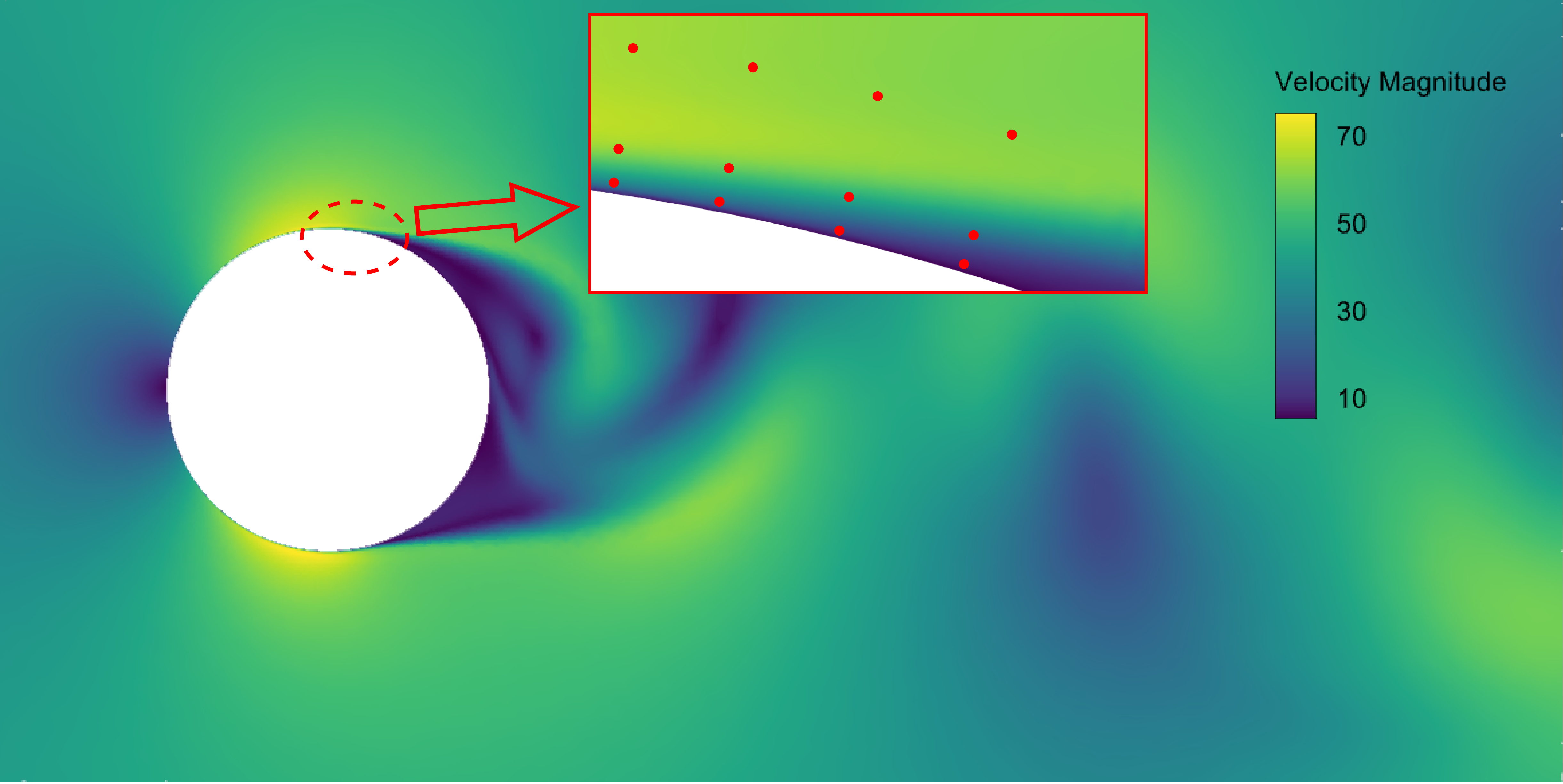}
    \caption{Schematic diagram of three layers of sensors placed around the surface of the cylinder, ensuring that the first layer of sensors is placed within the boundary layer.}
    \label{fig:6}
\end{figure}

Zooming in on the figure reveals that we have placed three layers of probes around the cylinder, marked with red dots, as shown in Fig.~\ref{fig:6}. The first layer is placed at a height of $0.1mm$ from the cylinder surface and can be used to measure airflow data over the surface of the cylinder; the second and third layers are placed at $1mm$ and $2.5mm$, respectively. It is crucial to ensure that the first layer of probes is within the boundary layer, enabling us to gather information about the flow separation points. The second and third layers of the probes aim to obtain information on pressure and velocity gradients, as well as information on the effects of flow separation on the wall surface, allowing the agent to learn and determine whether flow separation is exacerbated under the influence of the jets. In simpler terms, we wish the fluid to flow closely along the surface of the cylinder. Each layer consists of $128$ probes, each probe capable of simultaneously capturing two types of information: velocity and pressure. 

\begin{figure*}
    \centering
    \includegraphics[width=\linewidth]{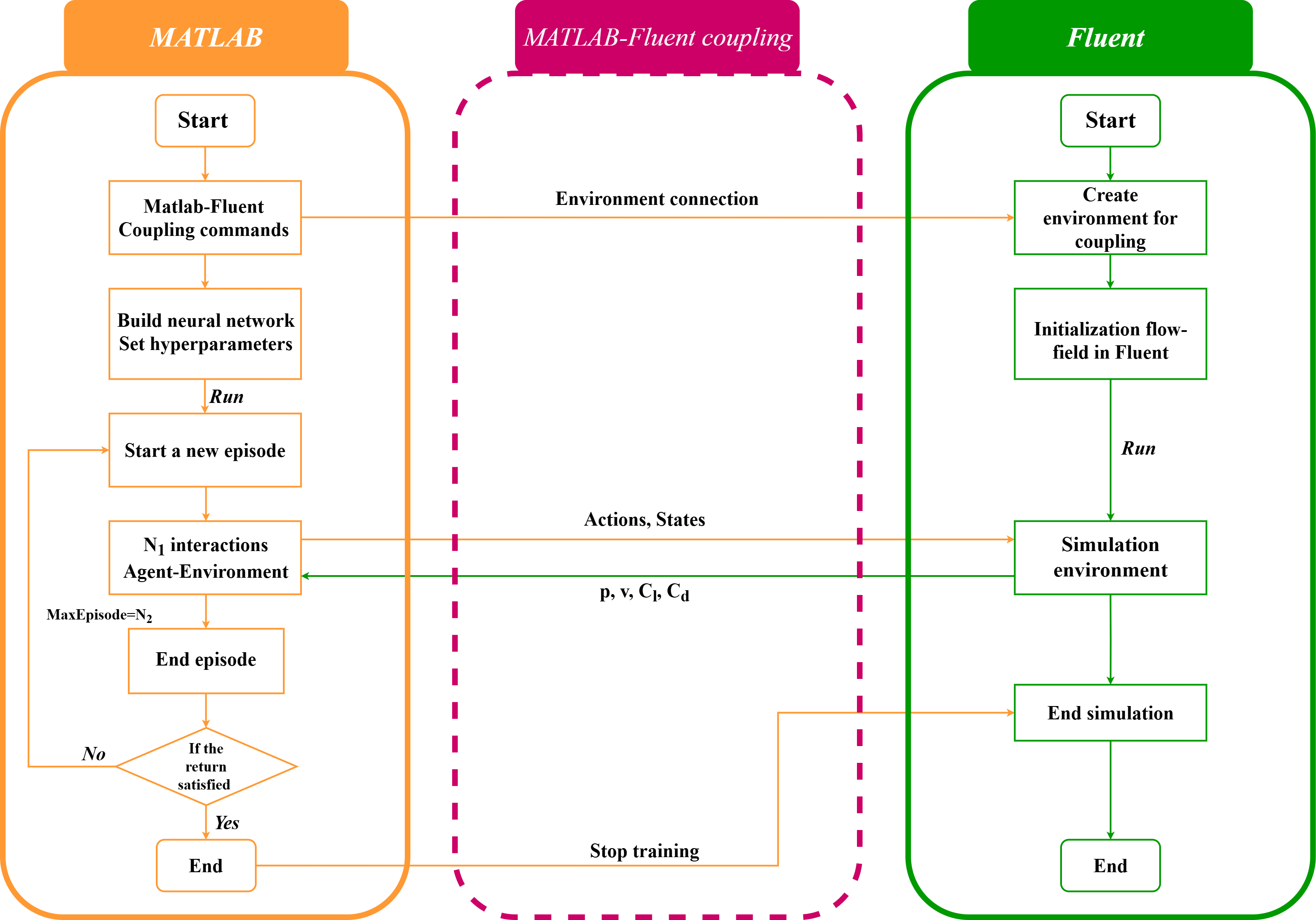}
    \caption{Flow chart of the Matlab-Fluent coupling with training process and agent-environment interactions.}
    \label{fig:7}
\end{figure*}

To facilitate the interaction between the agent and the environment, we establish a framework as shown in Fig.~\ref{fig:7}. Using a Text User Interface (TUI), a dialogue is established between Matlab and Fluent, with Matlab constructing the PPO agent. Matlab outputs the control variable, which is the maximum value of jet velocity, and this is fed into Fluent via the TUI, followed by the complete input of velocity inlet through the velocity profile equation. Once all boundary conditions have been entered, Fluent runs a simulation for one time step, which includes $n$ iterative steps. A complete interaction between Matlab and Fluent of $N_1$ times constitutes one episode, with $N_2$ being the total number of episodes in a training session. In our study, rather than resetting the flow field to its initial state at the end of an episode, the flow field is simulated for several time steps in an environment with only an inlet and no jets, allowing the flow field to return to an uncontrolled state, effectively conducting a random initialization. 

\section{Results and Discussions}\label{sec: result_and_discuss}

\begin{figure}[h]
    \centering
    \includegraphics[width=0.7\linewidth]{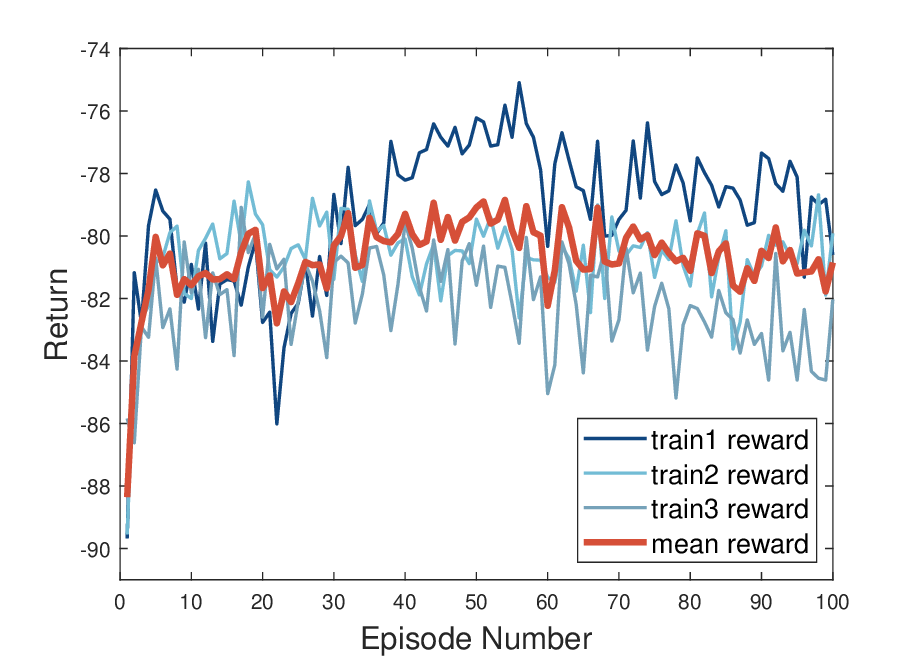}
    \caption{Training curve of PPO agent with three independent trainings.}
    \label{fig:8}
\end{figure}

\begin{figure*}[h]
\subfigure[]{\includegraphics[width = .49\linewidth]{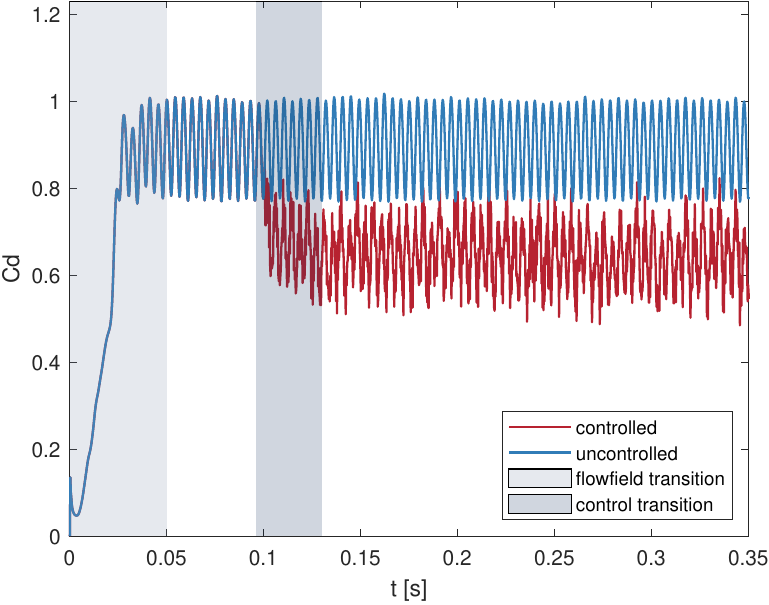}}\hfill
\subfigure[]{\includegraphics[width = .49\linewidth]{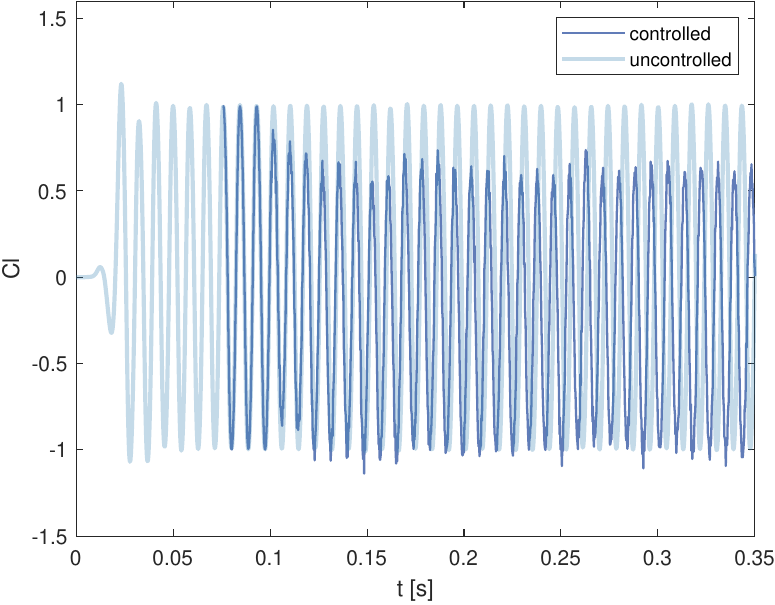}}
\caption{\label{fig:9}Comparison of the drag coefficient and lift coefficient of the cylinder under agent control versus uncontrolled conditions.}
\end{figure*}

As discussed previously, training an agent for control in high Reynolds-number flows is a time-consuming and uncertain endeavor. Each complete training session encompasses 100 episodes, with each episode containing 100 interactions, and each training session takes approximately 120 hours (hardware:i7-12700H @2,3GHz RAM 64GB). The return for each episode is the result of the weighted summation of the rewards over an entire trajectory. To demonstrate the reproducibility of the training, three independent training sessions were conducted, as shown in Fig.~\ref{fig:8}, and the returns of these sessions were averaged to obtain the mean return. It can be seen from the graph that the fluctuation in rewards during each training session is significant. Given the high time cost and the complexity of the environment, it is undesirable for the agent to remain in a local optimum during training. The agent should be more exploratory, since staying in a local optimum would result in a more stable and convergent return curve, but achieving a similarly effective control outcome would require exponentially more time. Thus, balancing training costs while obtaining a reliable control strategy becomes a challenging dilemma. We found that the mean return stabilizes around $-81$ after some fluctuations, indicating that individual training sessions are repeatable. Although we cannot conclude that this is the optimal control strategy (as there is no definitive criterion for determining an optimal control strategy), it suggests that finding a better control strategy than the current training results is extremely difficult. In this case, we need to verify the repeatability of the control strategy of the trained agent, in other words, whether agents with similar return values obtained from different trainings have similar control effects. We also need to assess whether the agent's control strategy meets the requirements at the control level, such as quick response speed and robustness. These aspects will be discussed further.

In the coefficient of drag graph, we disregard the data from the flow field transition phase and focus solely on the fully developed flow field starting from $0.05s$ onward, as shown in Fig.~\ref{fig:9}(a). This graph compares the drag coefficient of the cylinder without control to that under agent control after training completion. The control transition area, highlighted in a darker shade of gray, represents the process through which the flow field transitions from an uncontrolled state to a stable controlled state. The control strategy is observed to respond in approximately 0.03 seconds, as shown in Fig.~\ref{fig:9}(a), quickly transitioning the flow field to a controlled state with a significant reduction of $29\%$ in the drag coefficient, demonstrating the practical importance of this control strategy. In the coefficient of lift graph, it is evident that the amplitude of the oscillation is reduced. However, an asymmetric phenomenon in the vertical resultant force is also noticeable compared to the uncontrolled condition, as shown in Fig.~\ref{fig:9}(b). This implies that the symmetry of the agent's policy under this flow characteristic can be further improved based on specific requirements. When comparing the drag and lift curve graphs, it is not difficult to notice that the oscillation amplitude of the lift predominates.

\begin{figure*}

\subfigure[]{\includegraphics[width = .48\linewidth]{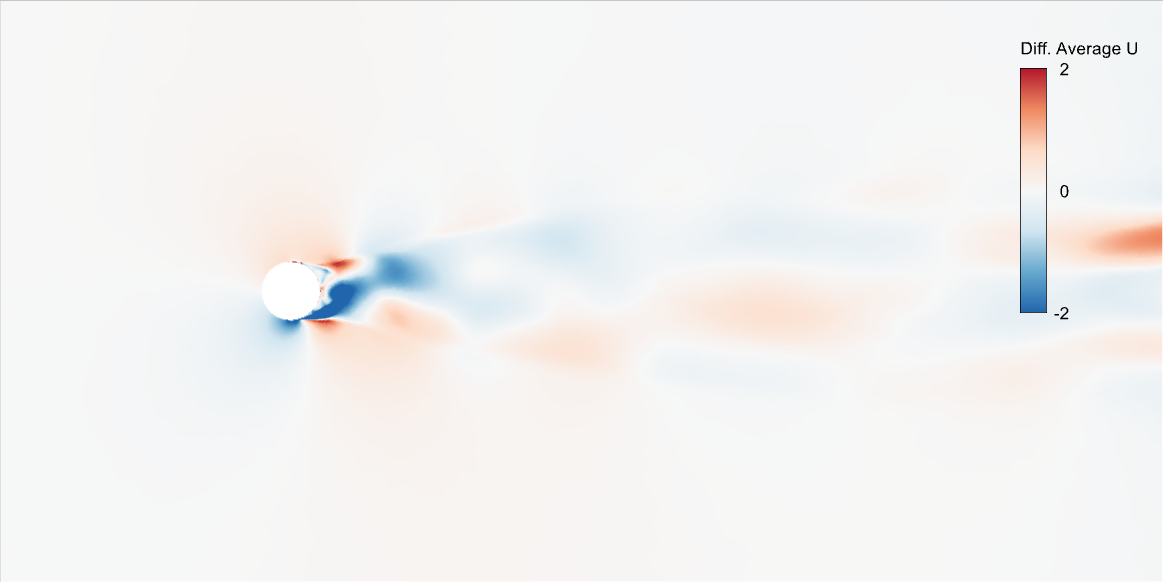}}
\subfigure[]{\includegraphics[width = .48\linewidth]{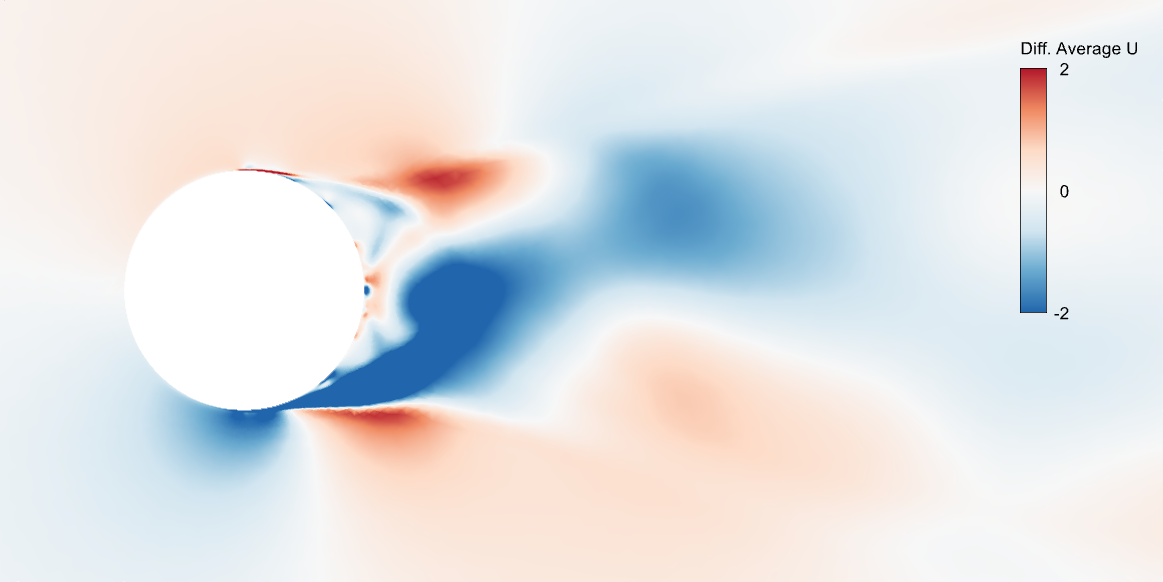}}
\\
\subfigure[]{\includegraphics[width = .48\linewidth]{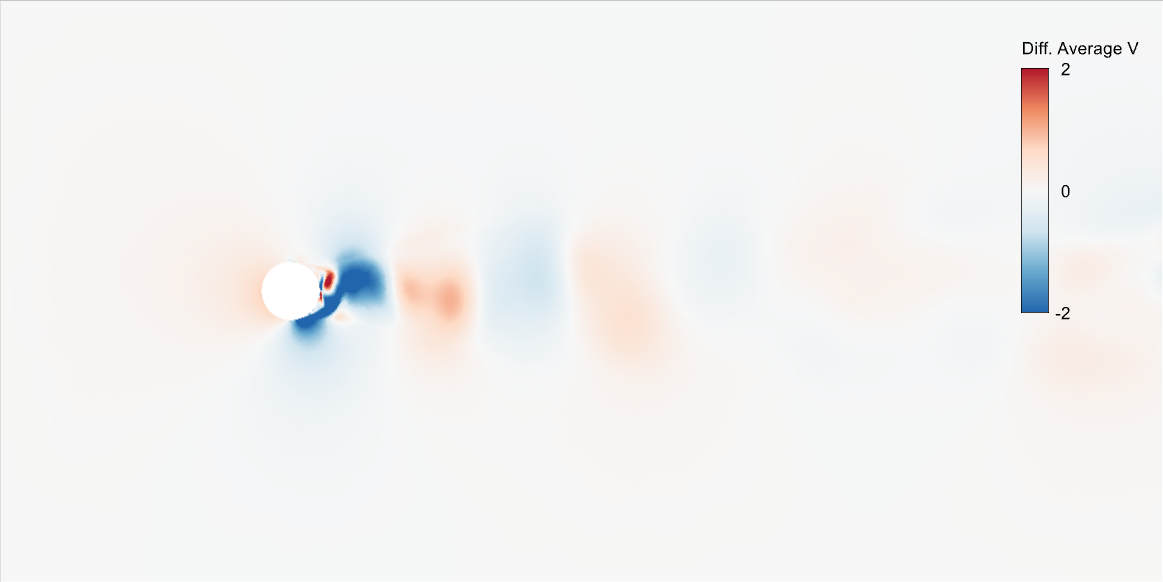}}
\subfigure[]{\includegraphics[width = .48\linewidth]{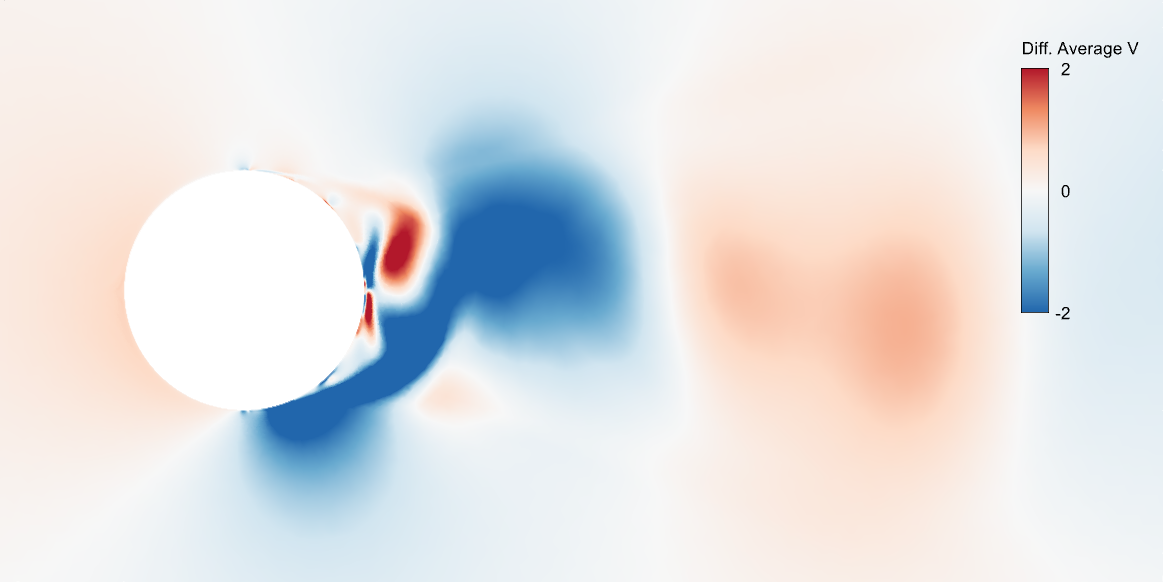}}
\\
\subfigure[]{\includegraphics[width = .48\linewidth]{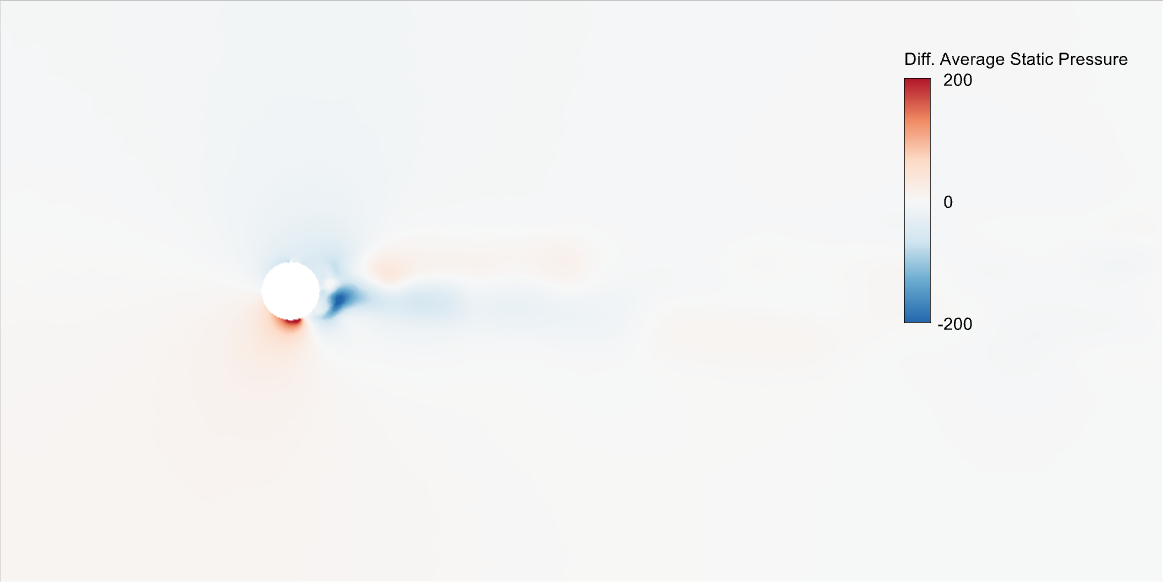}}
\subfigure[]{\includegraphics[width = .48\linewidth]{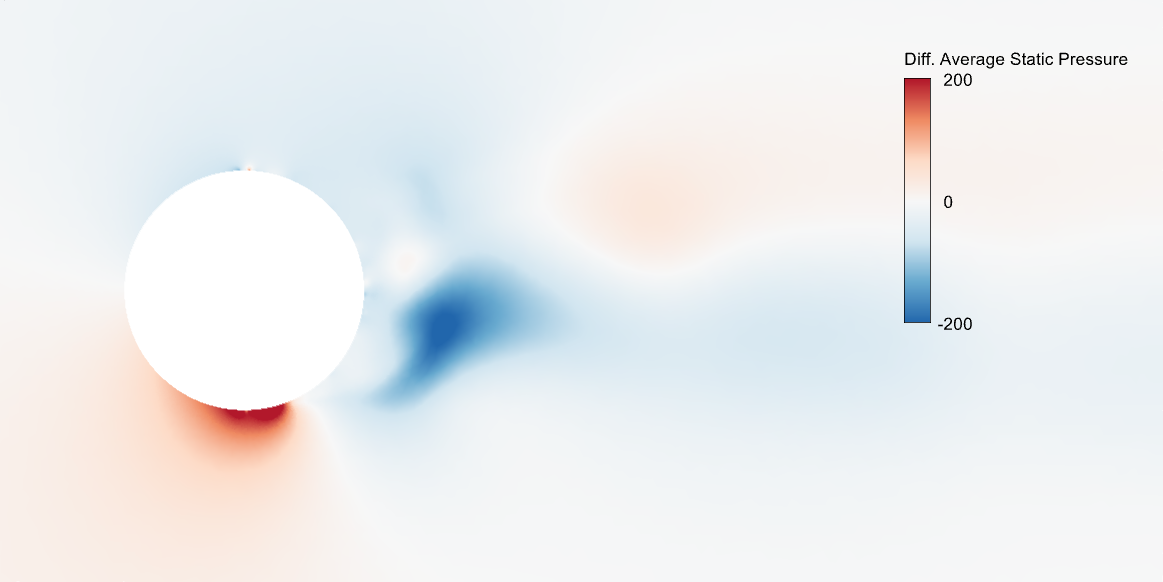}}
\\
\subfigure[]{\includegraphics[width = .48\linewidth]{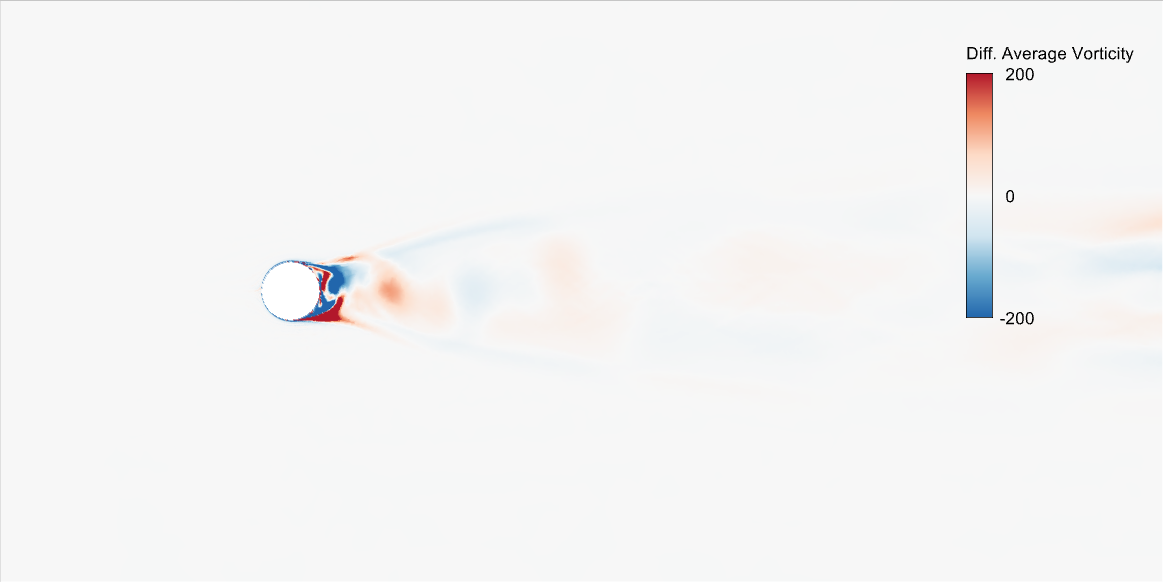}}
\subfigure[]{\includegraphics[width = .48\linewidth]{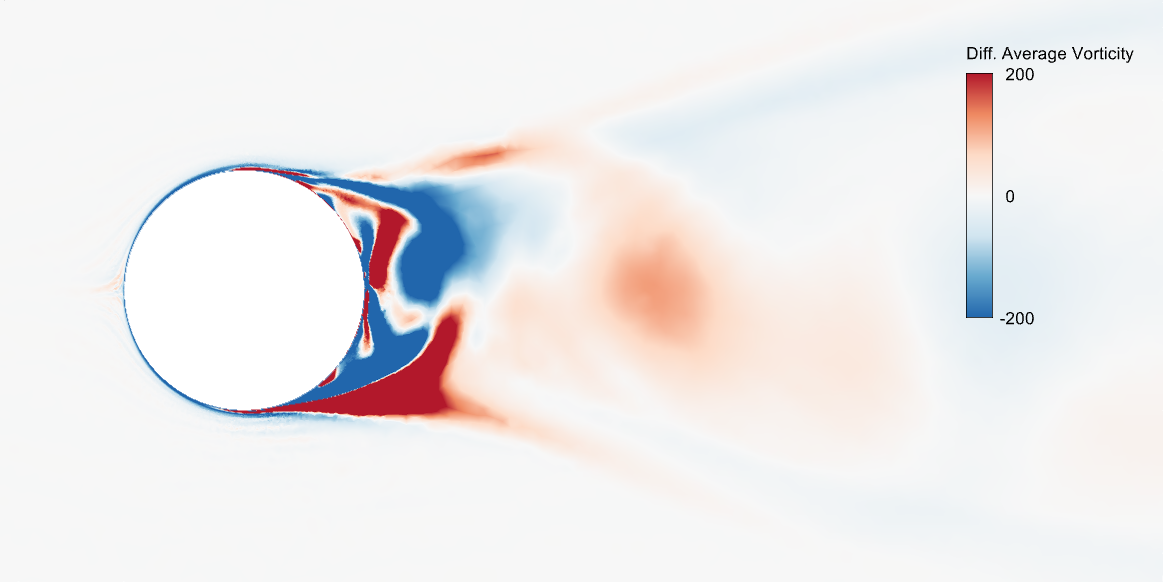}}

\caption{\label{fig:10} Contour of averaged difference of streamwise velocity (a), averaged difference of transverse velocity (b), averaged difference of static pressure (e) and averaged difference of vorticity (g) over 100 oscillation periods; the contours on the right column (b), (d), (f), (h) are enlarged views.}
\end{figure*}

From a statistical perspective, the flow field data are averaged over time to obtain the result of the difference between the flow field without control and the flow field under the control of a trained agent. In particular, the average is performed after simulating $100$ periods of oscillation. By calculating the difference, we can more intuitively observe the changes in the flow field caused by the jet, as the maximum speed of the jet is only $2m/s$, while the maximum speed around the cylinder can reach $70m/s$. The wake of the flow field is observed to be relatively symmetric, indicating that the flow field is statistically stable under control. However, slight asymmetry can be noticed around the cylinder through the mean streamwise velocity, the mean transverse velocity, and the mean static pressure, with a velocity difference of approximately $2m/s$. It is important to note that the maximum speed of the jet is also controlled at $2m/s$, with the inlet flow speed at $40m/s$. This slight asymmetry is understandable compared to the results achieved by the control. Possibly, this asymmetry is due to the uncertainties in the flow and the training, which prevent the agent from quickly mastering the macroscopic periodic patterns of the flow. For future work, it could be considered to integrate algorithms that can recognize signal features to explore whether agents can quickly learn the macroscopic periodic patterns of the fluid, and to investigate whether this method could accelerate agent training or stabilize the training at a better control effect. Additionally, the symmetry in the mean vorticity diagram is relatively good, indicating that the jet has a minimal impact on vorticity. 

\begin{figure}
    \centering
    \includegraphics[width=0.8\linewidth]{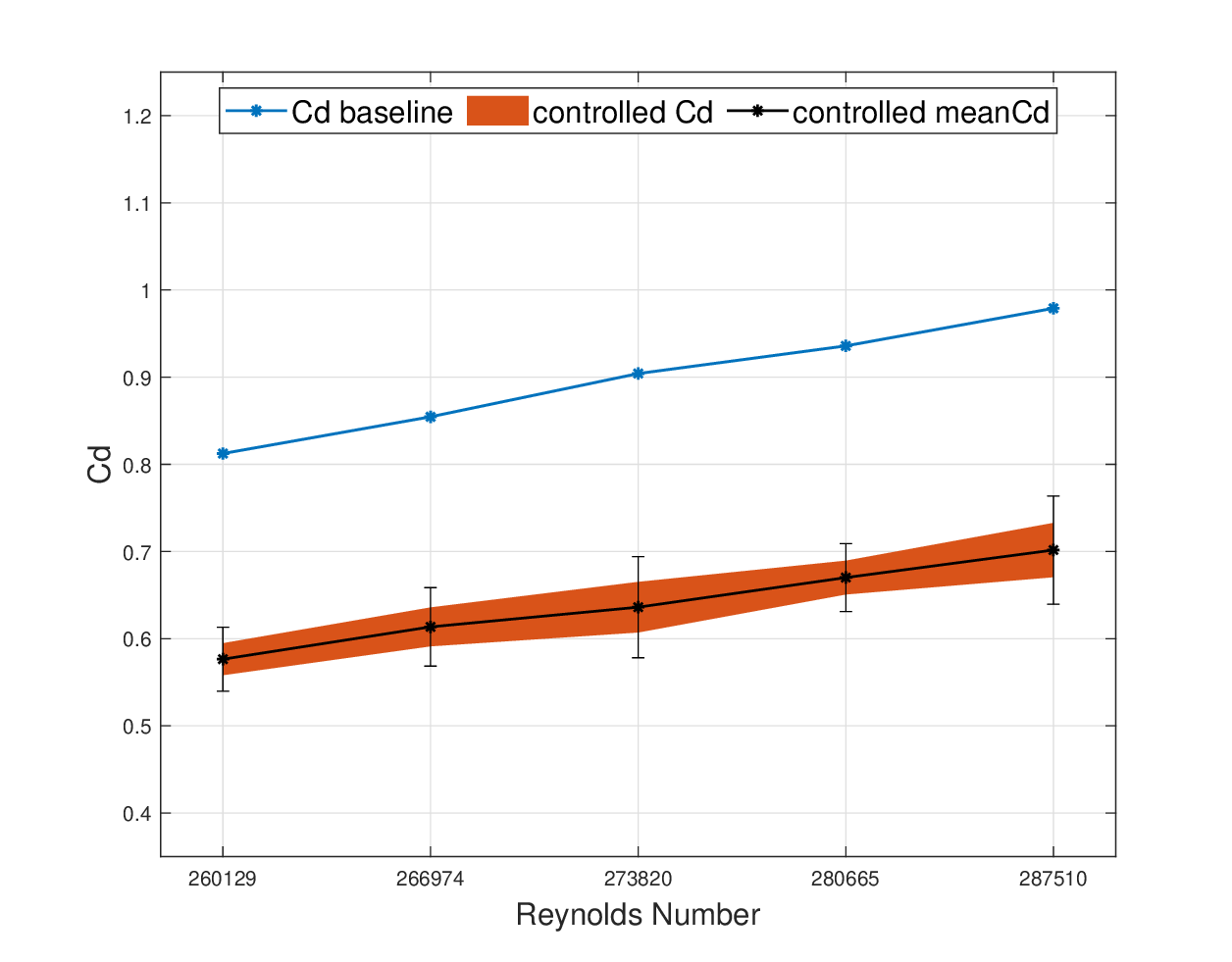}
    \caption{Comparison of the coefficient of drag before and after control under a range of Reynolds number. The blue line is the baseline without control and the black line is the coefficient of drag averaged by five independent repeated control tests. The error bars denote the difference between the maximum and minimum coefficient of drag among these five tests. The orange area represents the standard deviation range.}
    \label{fig:11}
\end{figure}

To verify the repeatability of the control strategy, we selected an agent trained from one of the three independent training sessions conducted after the mean reward curve stabilized. This agent was subjected to five repeated control tests, each lasting $100$ periods of oscillation, during which the drag coefficient was calculated. This process was also repeated with variations in Reynolds numbers to assess the robustness of the control strategy, as shown in Fig.~\ref{fig:11}. The blue curve represents the variation in the drag coefficient under different Reynolds numbers without control. The orange area indicates the range of variance in the drag coefficient after five tests, while the black curve represents the average drag coefficient under different Reynolds numbers after testing. The error bars show the difference between the maximum and minimum drag coefficients obtained in five tests at each Reynolds number. The results of five tests with different Reynolds numbers show that the drag under agent control varies within a small range, proving that the control strategy is repeatable and stable. Comparison of drag coefficients before and after control under different Reynolds numbers shows that the control strategy possesses a certain degree of robustness.

\begin{figure*}
\includegraphics[width=\linewidth]{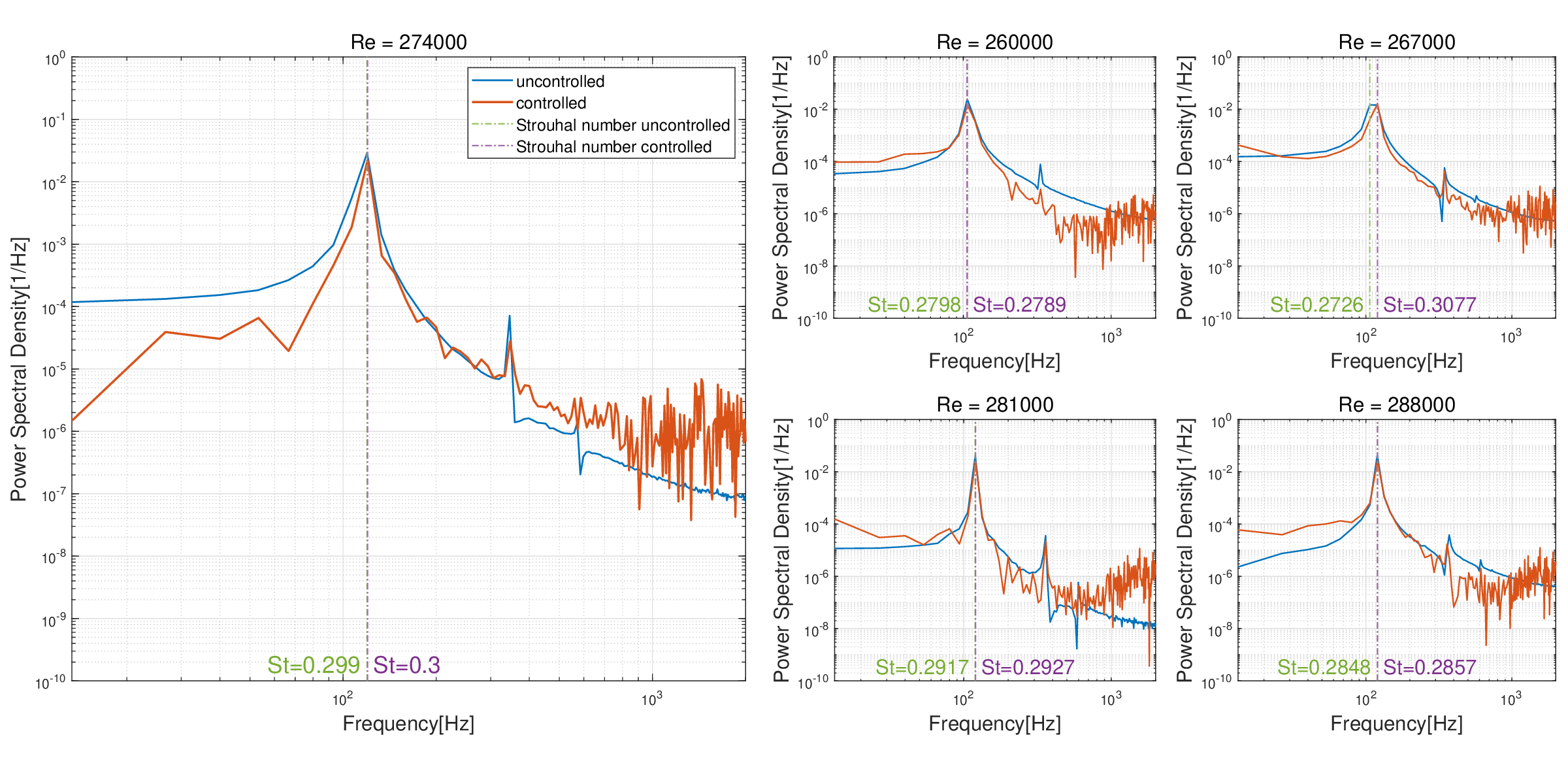}
\caption{\label{fig:12}Diagram of the comparison of Power Spectral Density (PSD) before and after control at different Reynolds number, marked with Strouhal number.}
\end{figure*}

\begin{figure}
\subfigure[]{\includegraphics[width = .41\linewidth]{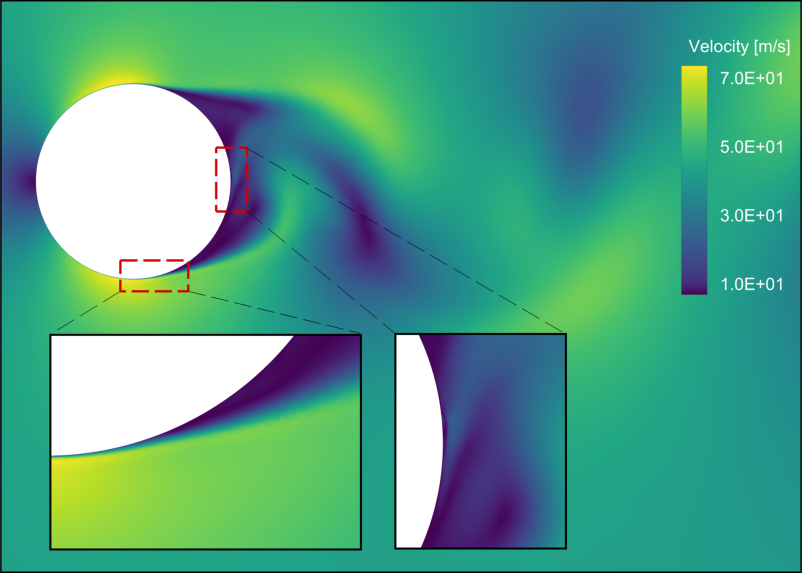}}
\subfigure[]{\includegraphics[width = .41\linewidth]{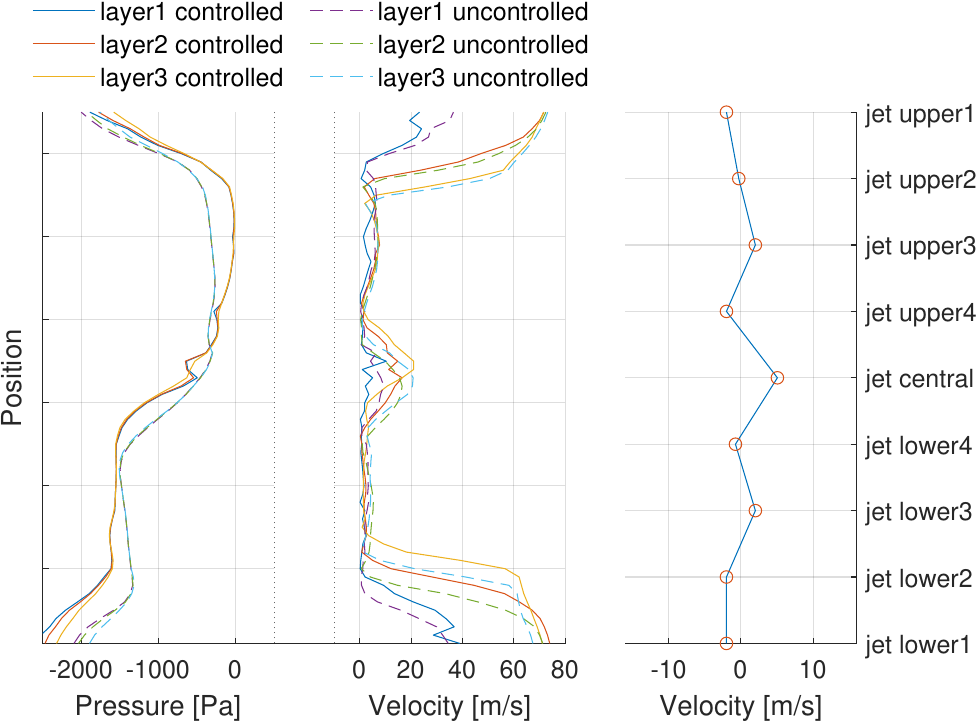}}
\\
\subfigure[]{\includegraphics[width = .41\linewidth]{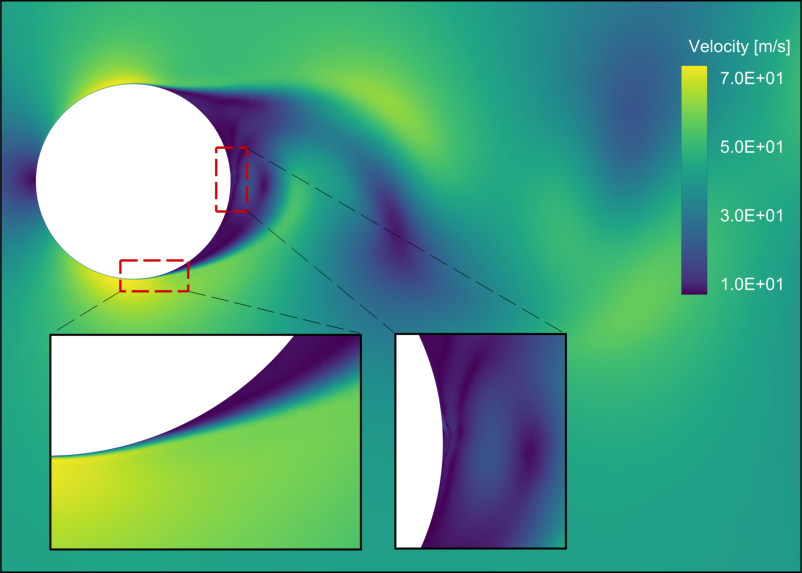}}
\subfigure[]{\includegraphics[width = .41\linewidth]{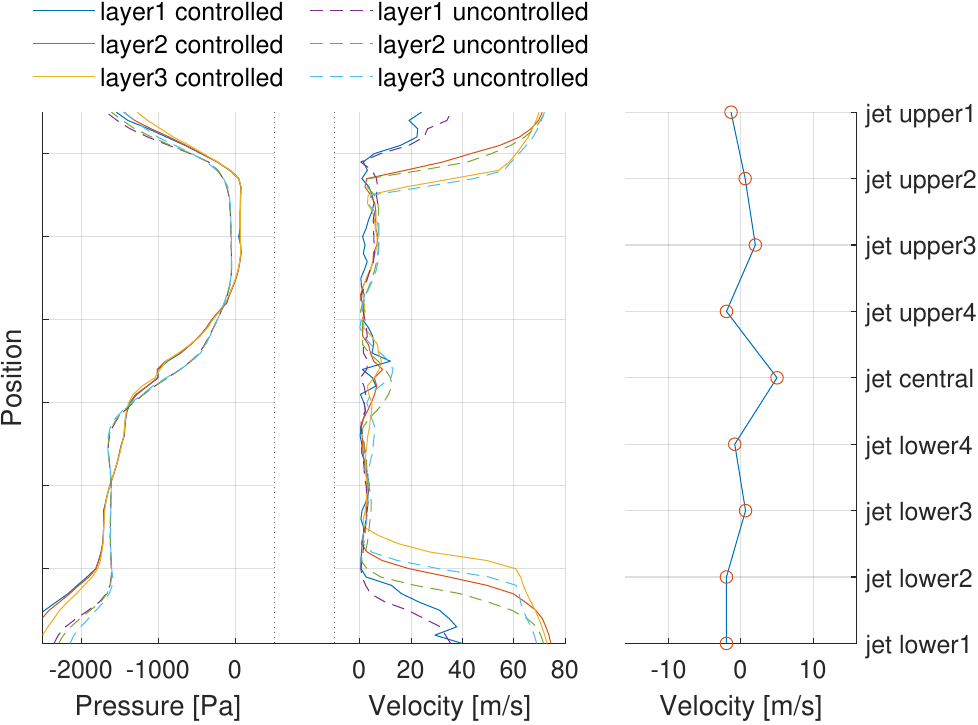}}
\\
\subfigure[]{\includegraphics[width = .41\linewidth]{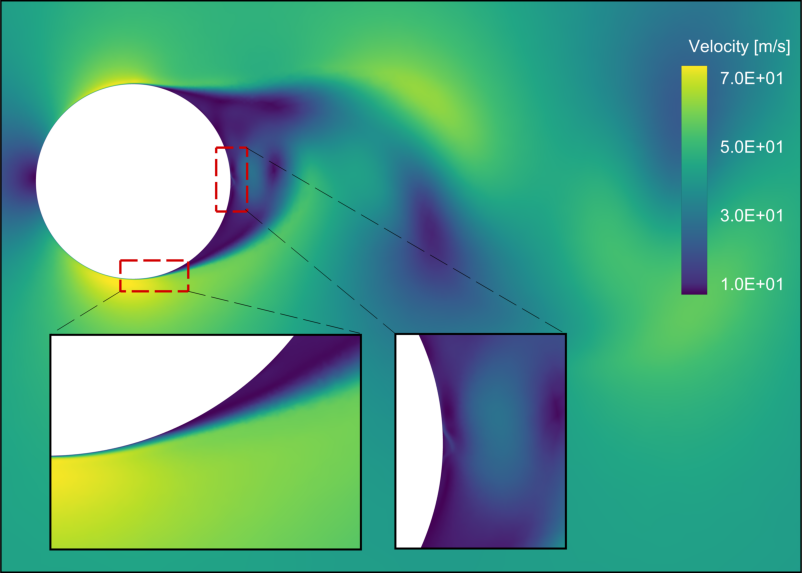}}
\subfigure[]{\includegraphics[width = .41\linewidth]{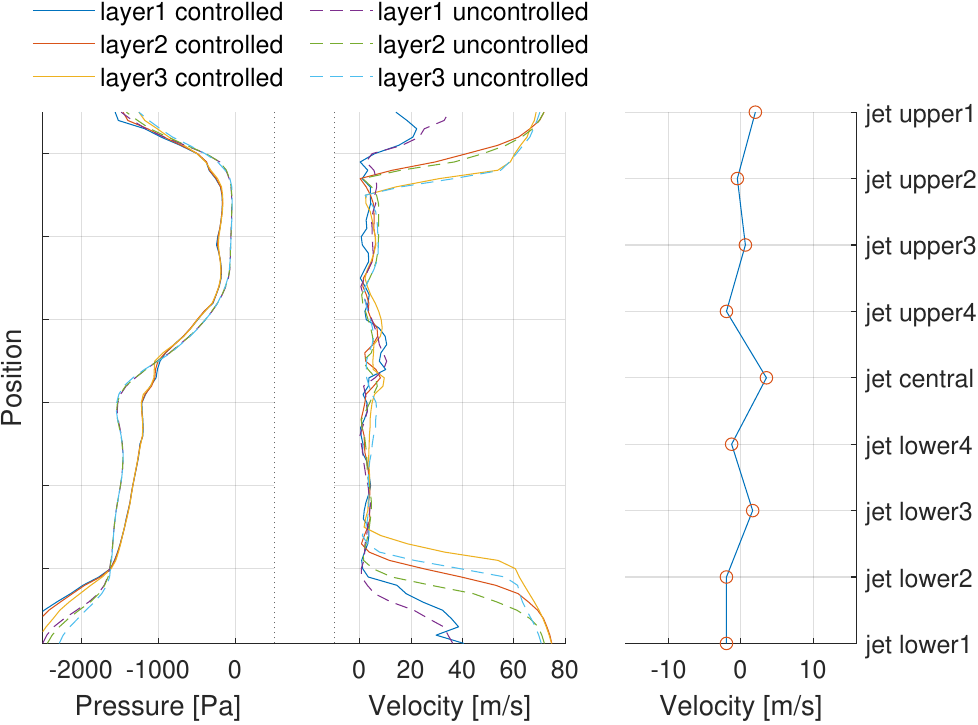}}
\\
\subfigure[]{\includegraphics[width = .41\linewidth]{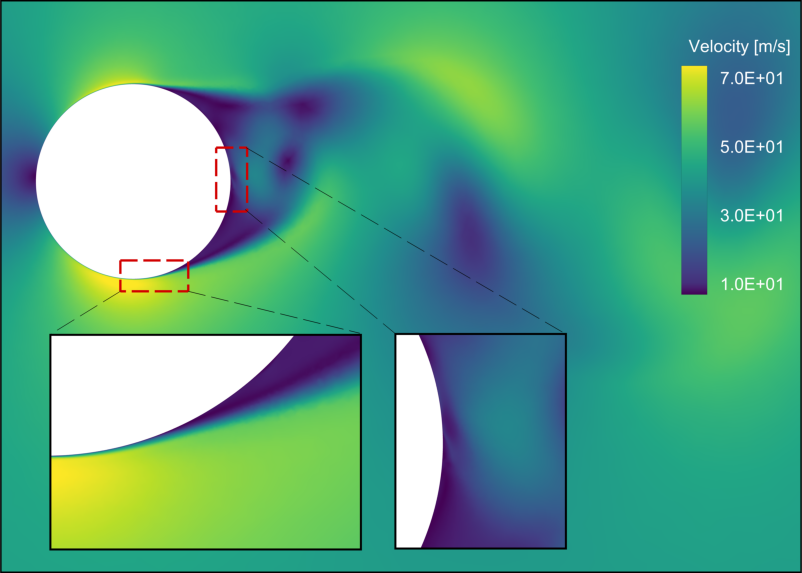}}
\subfigure[]{\includegraphics[width = .41\linewidth]{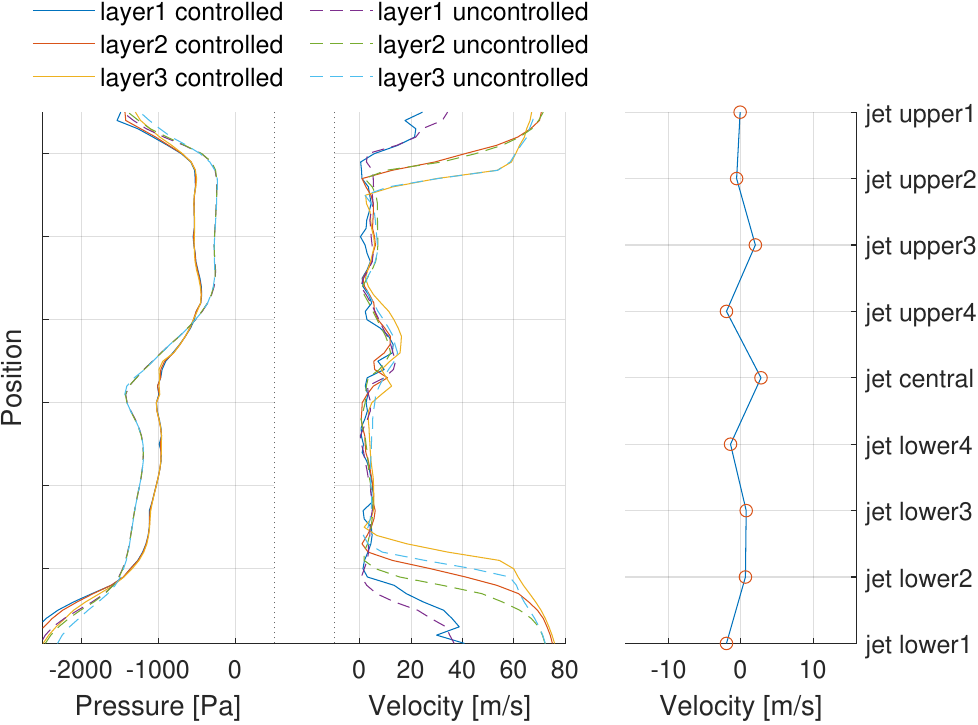}}
\end{figure}

\begin{figure}
\subfigure[]{\includegraphics[width = .41\linewidth]{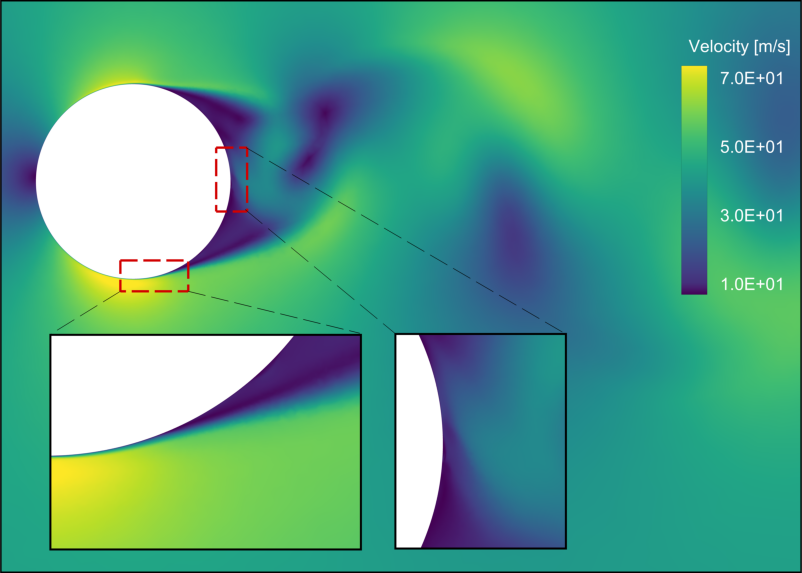}}
\subfigure[]{\includegraphics[width = .41\linewidth]{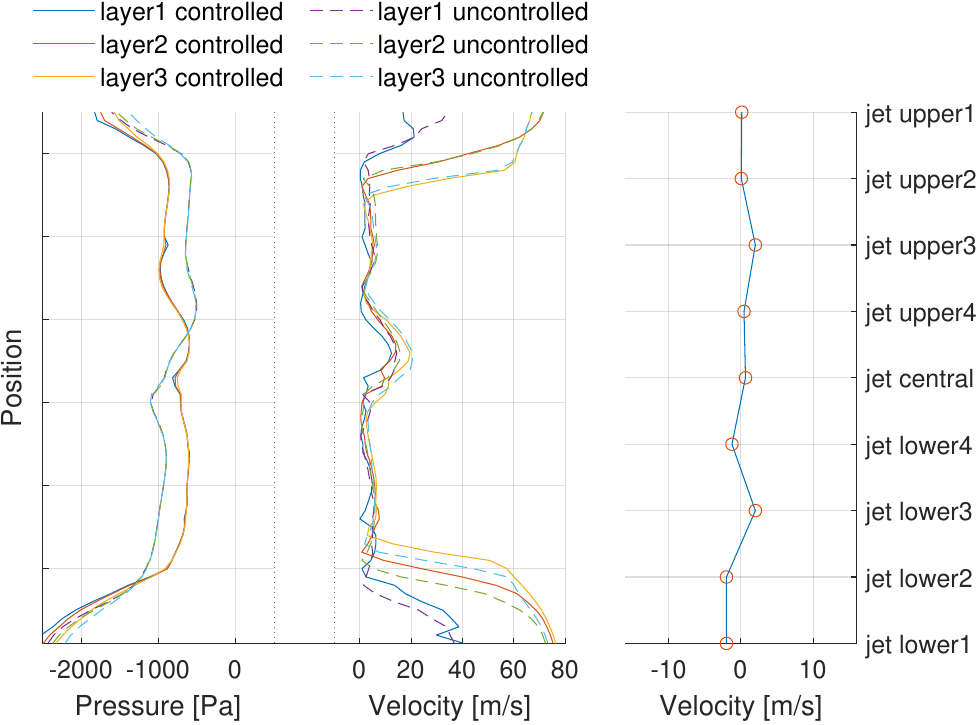}}
\\
\subfigure[]{\includegraphics[width = .41\linewidth]{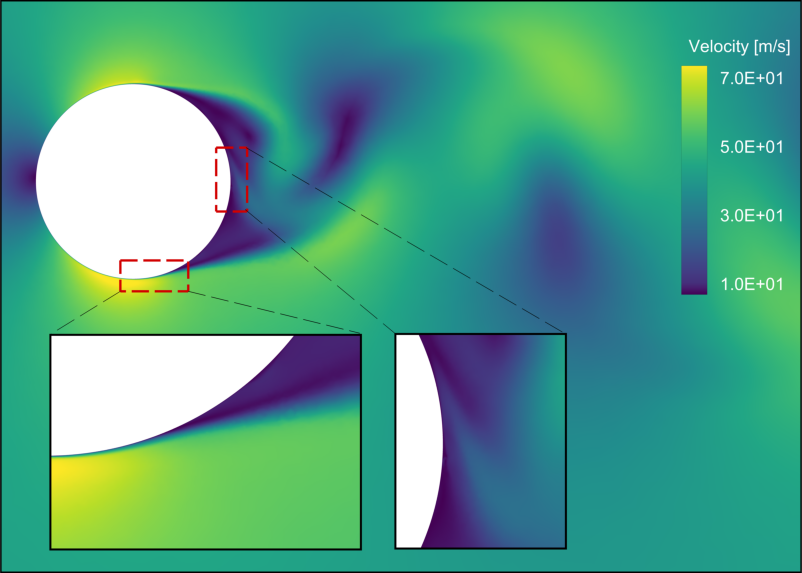}}
\subfigure[]{\includegraphics[width = .41\linewidth]{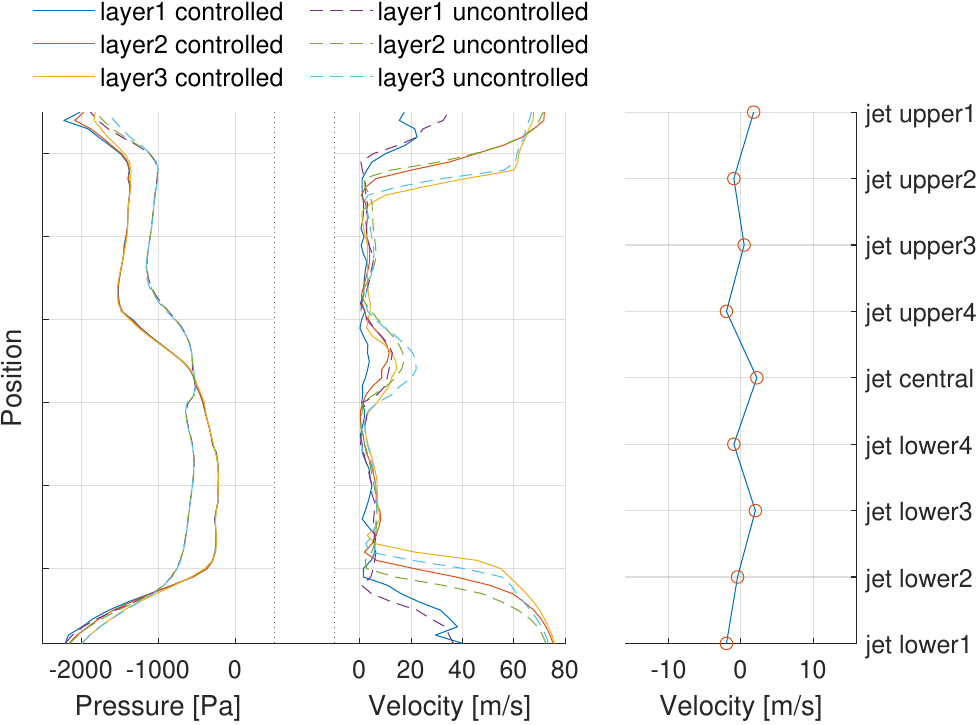}}
\caption{\label{fig:13}Visualization of the agent policy: Contour of velocity magnitude with zoomed in views on the left column: (a), (c), (e), (g), (i), (k). The comparison of the velocity and pressure detected by three layers of sensors on the right half of the cylinder corresponding the velocity of nine jets on the right column.}
\end{figure}

\begin{figure}
\subfigure[]{\includegraphics[width = .31\linewidth]{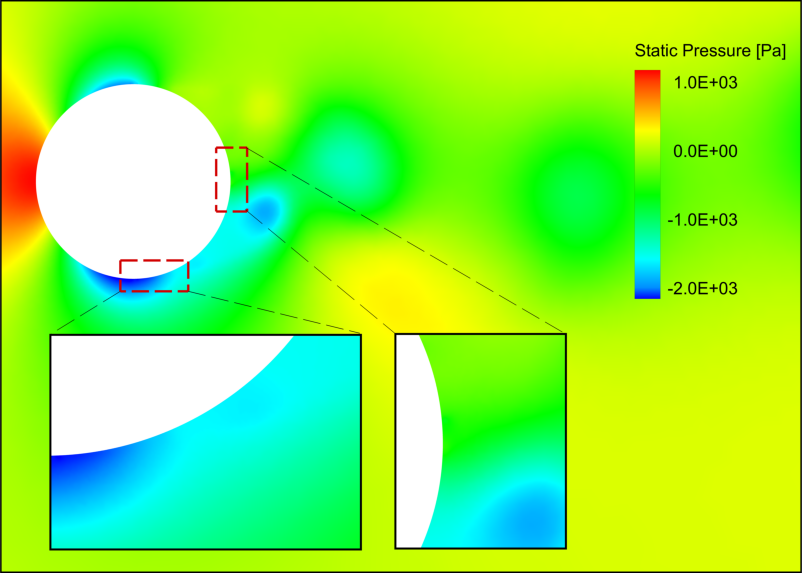}}
\subfigure[]{\includegraphics[width = .31\linewidth]{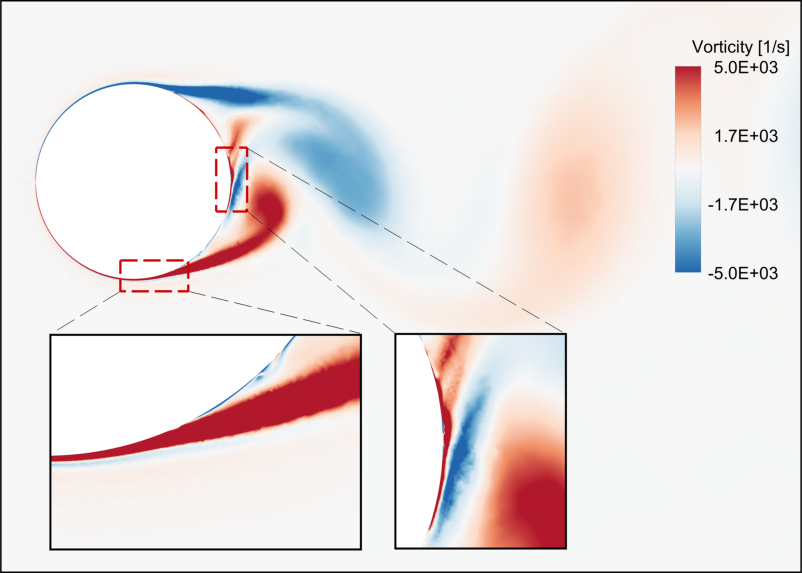}}
\subfigure[]{\includegraphics[width = .31\linewidth]{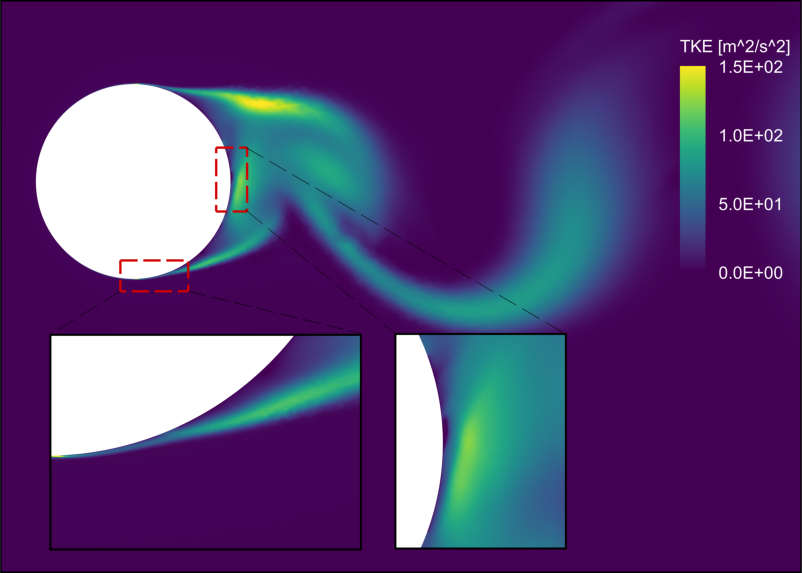}}
\\
\subfigure[]{\includegraphics[width = .31\linewidth]{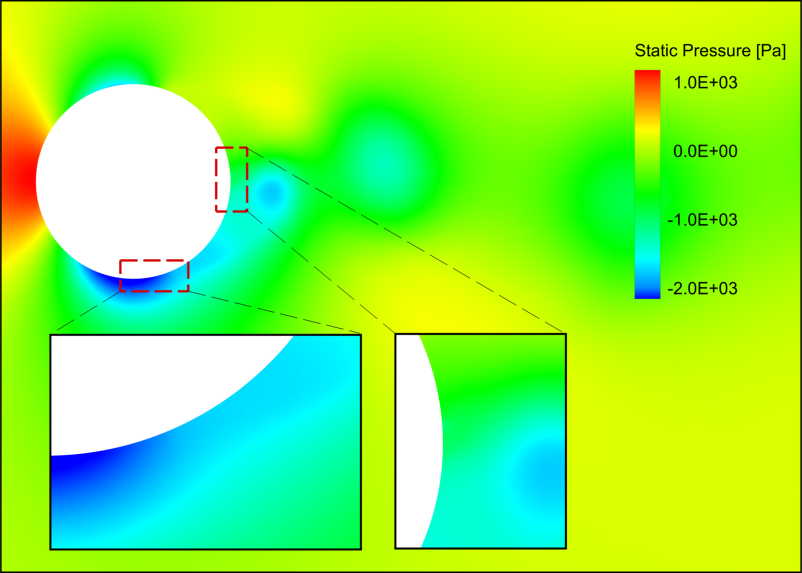}}
\subfigure[]{\includegraphics[width = .31\linewidth]{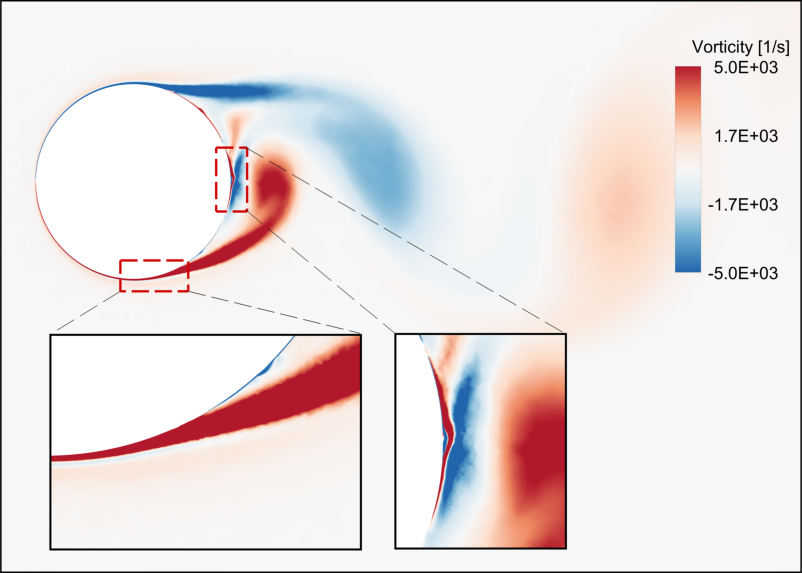}}
\subfigure[]{\includegraphics[width = .31\linewidth]{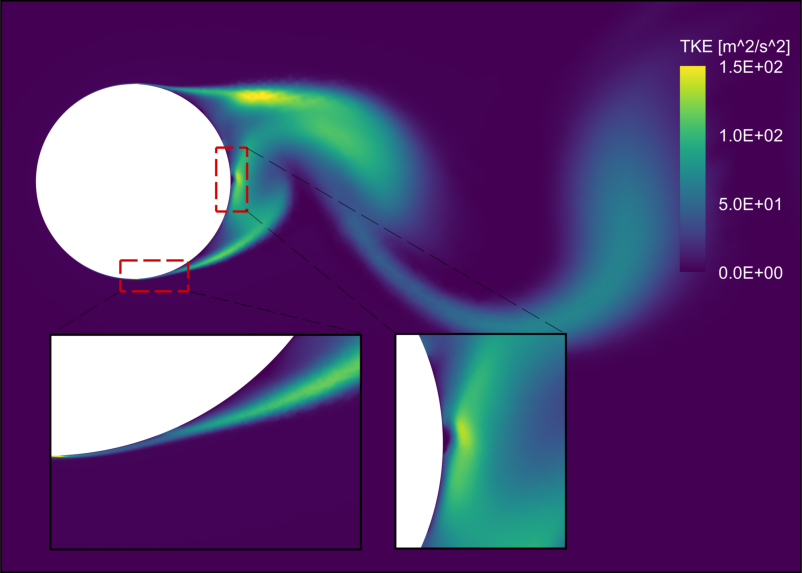}}
\\
\subfigure[]{\includegraphics[width = .31\linewidth]{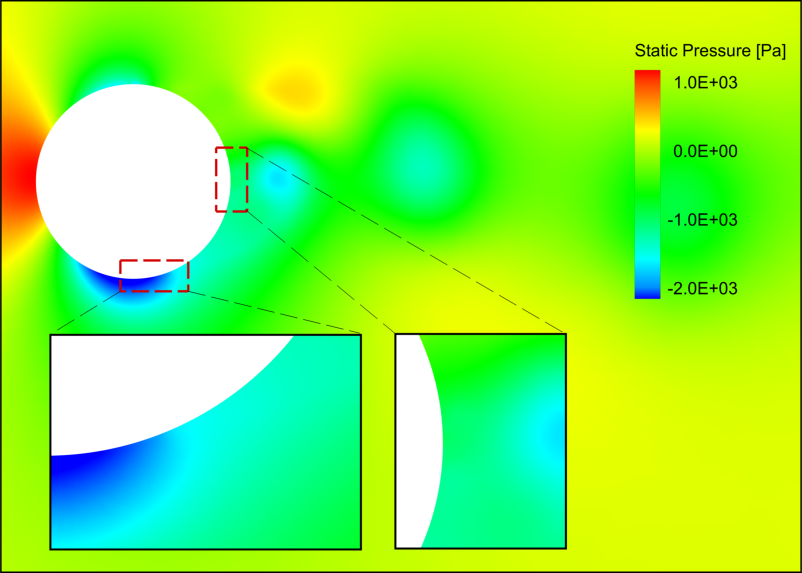}}
\subfigure[]{\includegraphics[width = .31\linewidth]{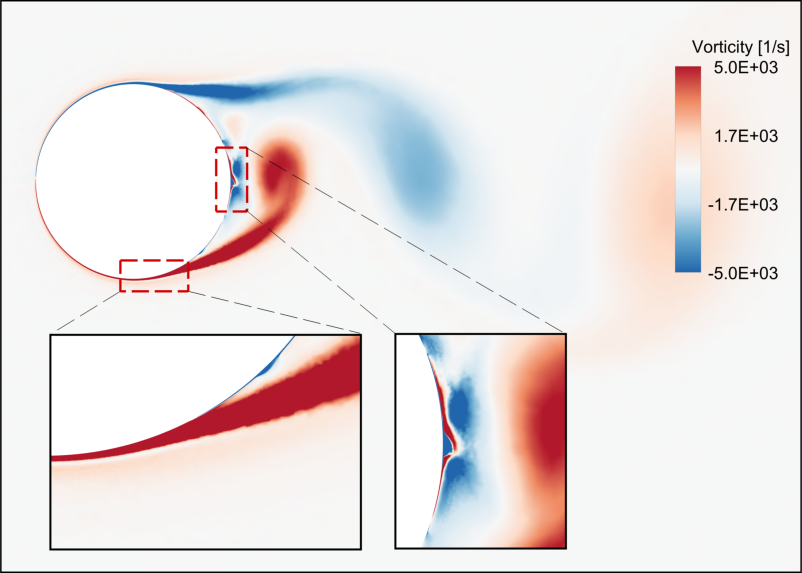}}
\subfigure[]{\includegraphics[width = .31\linewidth]{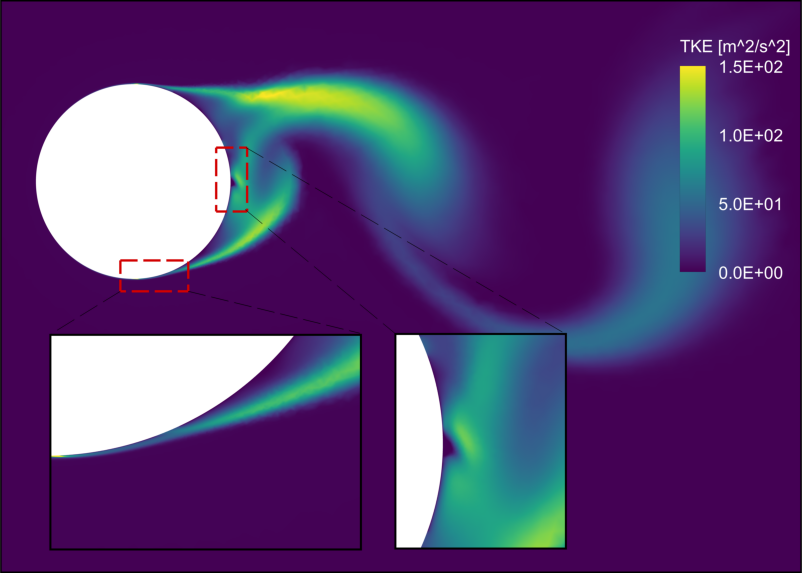}}
\\
\subfigure[]{\includegraphics[width = .31\linewidth]{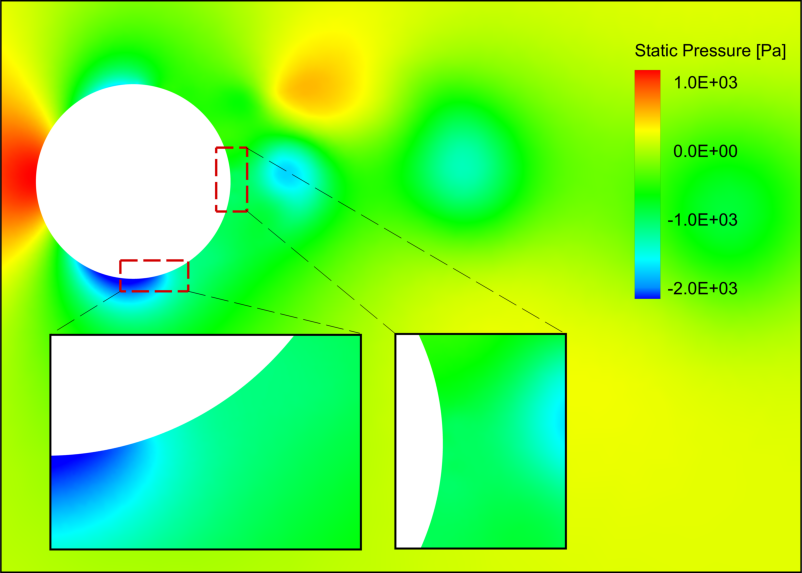}}
\subfigure[]{\includegraphics[width = .31\linewidth]{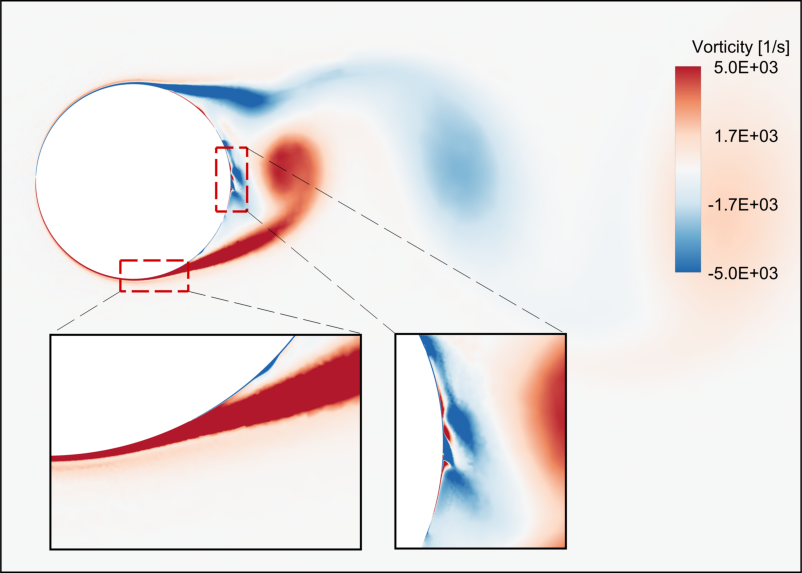}}
\subfigure[]{\includegraphics[width = .31\linewidth]{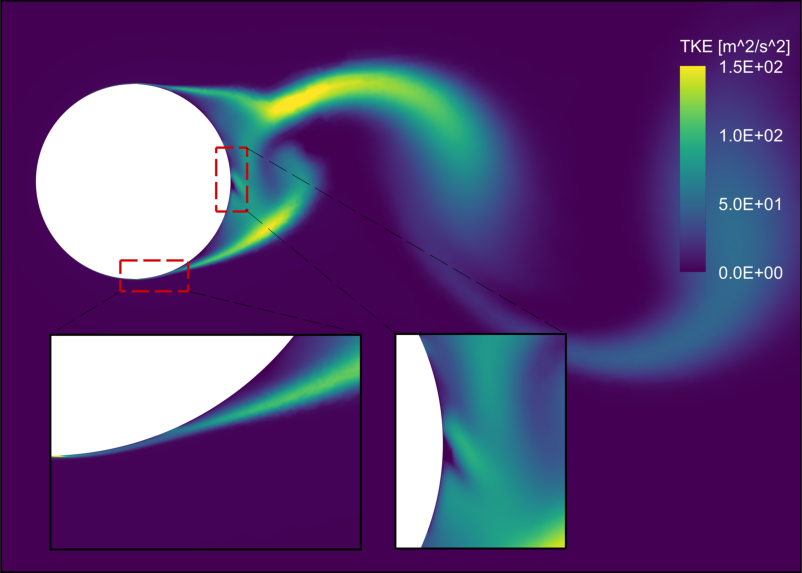}}
\\
\subfigure[]{\includegraphics[width = .31\linewidth]{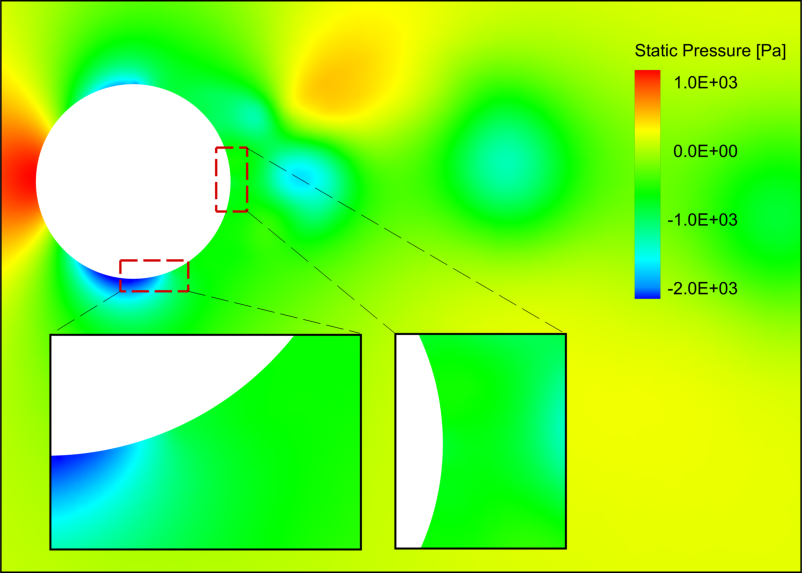}}
\subfigure[]{\includegraphics[width = .31\linewidth]{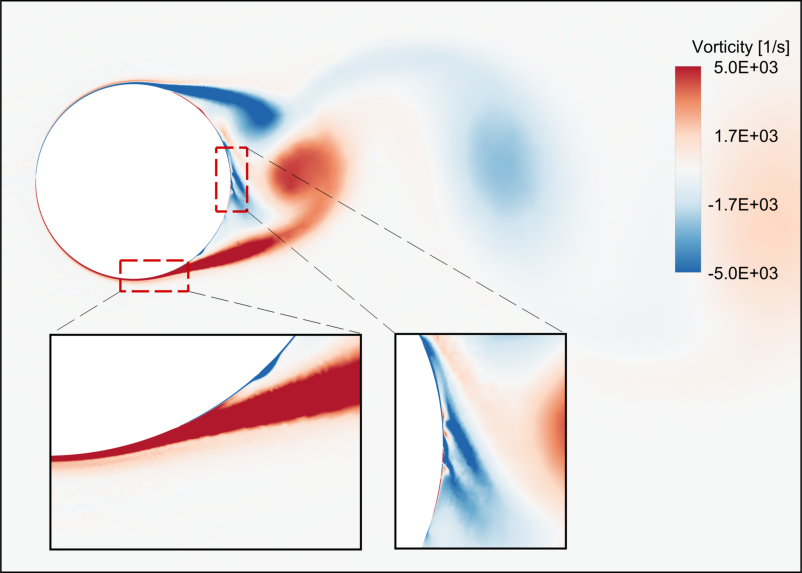}}
\subfigure[]{\includegraphics[width = .31\linewidth]{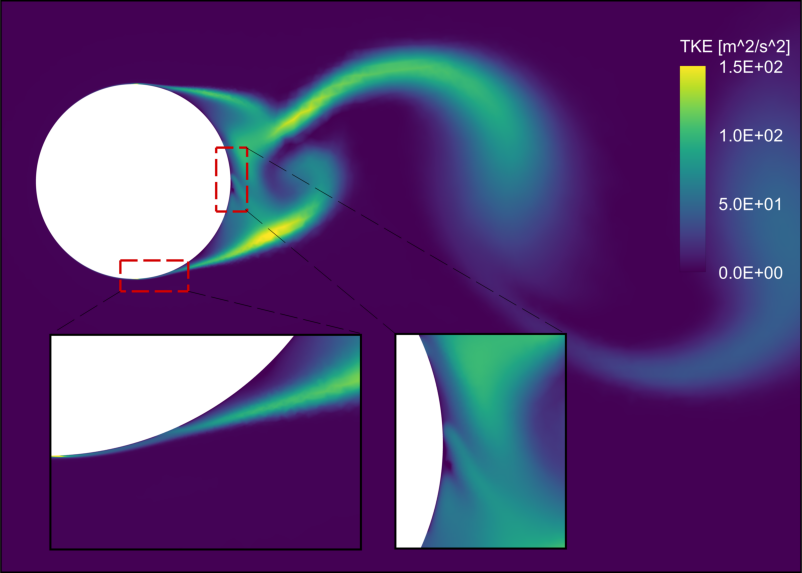}}

\end{figure}

\begin{figure}
\subfigure[]{\includegraphics[width = .31\linewidth]{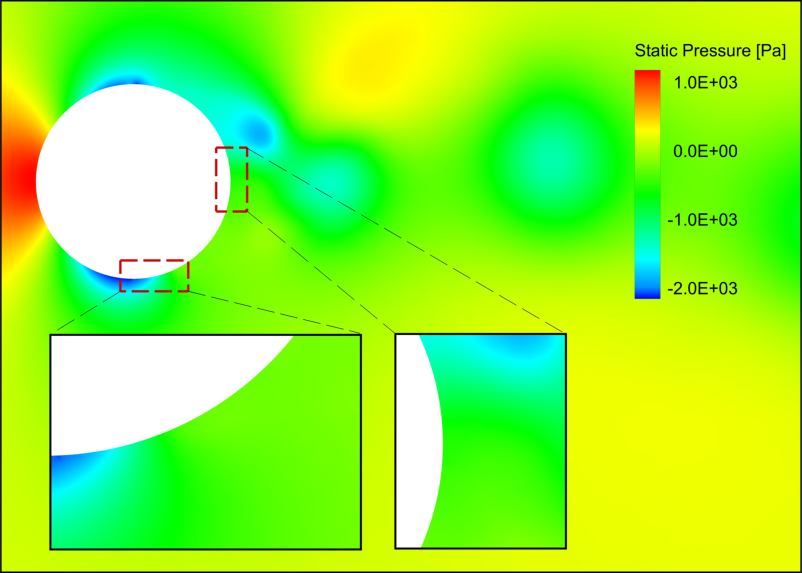}}
\subfigure[]{\includegraphics[width = .31\linewidth]{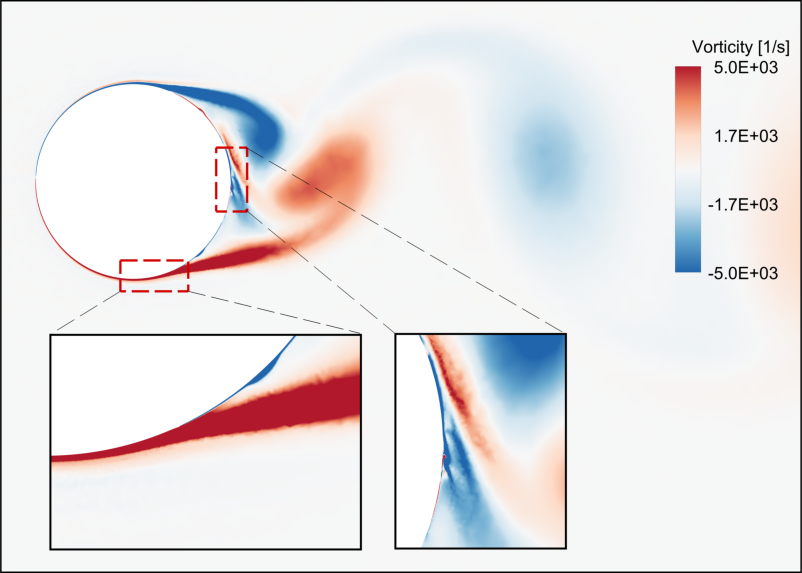}}
\subfigure[]{\includegraphics[width = .31\linewidth]{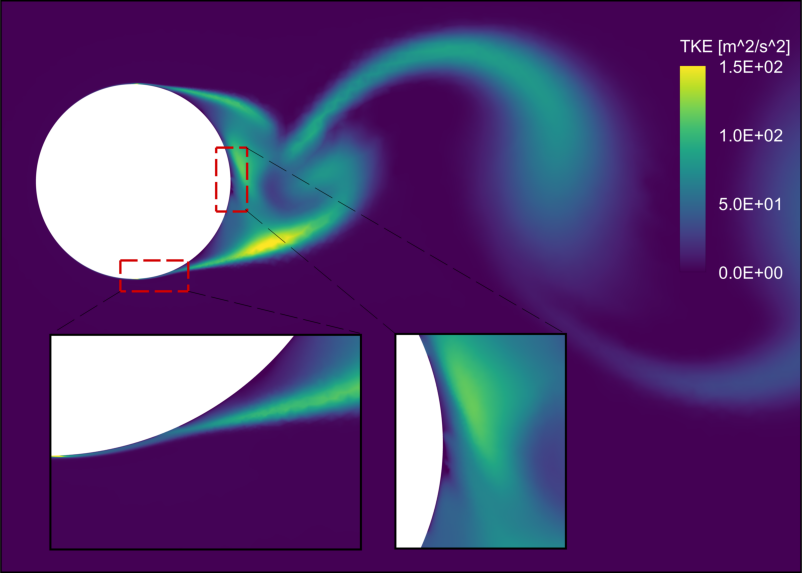}}
\caption{\label{fig:14} Comparison of the static pressure, vorticity and turbulent kinetic energy (TKE) at different times with time interval 0.001s.}
\end{figure} 

In tests carried out in various Reynolds numbers using an agent trained in Re = 244000, the robustness of the procedure was established. The study delved into the impact of the same agent on flow field structures under different flow conditions from a frequency domain perspective. The blue and orange curves in the graph represent the lift coefficient curves without control and under agent control, respectively, post Fast Fourier Transform (FFT). These curves illustrate the distribution of energy at different frequencies due to vertical oscillations acting on the cylinder, as shown in Fig.~\ref{fig:12}. The low-frequency portion of the graph indicates large-scale structures within the flow, while the high-frequency portion indicates smaller-scale oscillations. The vertical dashed lines in green and purple, respectively, denote the Strouhal numbers without control and under agent control, corresponding to the peaks of the two curves at the vortex-shedding frequency on the horizontal axis. This frequency is characteristic of the von Kármán vortex streak. Post-control, the vortex-shedding frequency exhibits varying degrees of change across different Reynolds numbers. In particular, only at Re = 274000 is there a significant suppression of low-frequency energy, indicating an effective mitigation of lift oscillations caused by large-scale vortices formed as the fluid passes the cylinder. However, this phenomenon was not as effectively observed under the control of this agent at other Reynolds numbers, suggesting that the agent's learning was specifically tuned to the frequency characteristics of the flow conditions at training. Due to the use of the RANS model, the impact of turbulent fluctuations across different scales in the frequency domain was not very pronounced, leading to relatively smoother curves. Nevertheless, a comparison of the orange curves across different Reynolds numbers reveals significant fluctuations in the energy in the high-frequency region post-control, primarily showing an increase. This suggests that a significant portion of the flow control impact targets small-scale flow structures. Within the studied range of Reynolds numbers, it is evident that high-frequency, small-scale structures in the flow field carry less energy. The blue curve further illustrates that as the scale of flow decreases, energy is transferred from large-scale to small-scale structures. The slope of the curve to the right of the vortex shedding frequency indicates that the control strategy does not significantly alter the distribution of energy across different scales or the inherent energy at the vortex shedding frequency.

The series of velocity contours at time levels $t_0 + 0.001n\ (n=0,1,2,3,4,5)$, as shown in Fig.~\ref{fig:13}, along with the velocity and pressure data extracted from three layers of sensors on the right-hand surface of the cylinder and the jet velocities of nine jets, visualize the control strategy or policy employed by the agent. The changes in velocity contour across different times and the fluctuations of pressure on the cylinder's surface elucidate the oscillatory process of vortex shedding. The enlarged images reveal the variations in the flow separation zone on the cylinder surface and how the shedding vortices reattach to the cylinder's back, impacting the pressure there. In Fig.~\ref{fig:13}(b), it is evident that under the influence of the central jet, the pressure at the back of the cylinder increases, resulting in a peak in the pressure curve.

The velocity diagram of the nine jets indicates that the central jet predominantly employs a jetting strategy, which results in the jet's reactive force directly propelling the cylinder and elevating the static pressure at its back, thereby reducing pressure drag. The velocities of jet upper1 and jet lower1 are minimal and exhibit little change, as increasing the velocity of these two jets would lead to premature flow separation and a larger separation zone, further decreasing the pressure at the cylinder's back and consequently increasing pressure drag. Therefore, the agent avoids employing a high-speed jetting strategy for these two jets. The other jets adjust their strategy periodically in response to the pulsations of the flow separation zone and the changes in the reattachment vortices, adopting a periodic strategy to optimize the control effect.

This series of contours reflect the dynamic changes in the flow field at time levels $t_0 + 0.001n\ (n=0,1,2,3,4,5)$, as shown in Fig.~\ref{fig:14}. By examining the zoomed-in views, the characteristics of the local flow can be observed. However, it is crucial to reiterate that, due to the use of the RANS model to simplify turbulence by time averaging for computational efficiency, not all details within the flow field are captured with high precision. However, by comparing the interaction between jets and vortex systems at different times across these three figures, we can reveal how jets achieve control objectives by influencing the flow field. The development of shedding vortices behind the cylinder can be seen through the changes in six sets of vorticity and turbulent kinetic energy contours, with alternating vortex streets forming above and below the cylinder. Zoomed-in views reveal a reattachment zone of shedding vortices on the cylinder's right-hand side (downstream), characterized by a lower static pressure and higher turbulent kinetic energy, indicating more disorderly flow. The vorticity contours also show that the flow in the reattachment zone exhibits periodicity, with these small structural vortex systems corresponding to the high-frequency portion mentioned previously in the power spectral density graph. Notably, a region near the central jet's position exhibits significantly low turbulent kinetic energy (identified in the vorticity contour between the red and blue vortices in the white area on the right zoomed-in view), resulting from small vortex systems generated around the jet. This is indicative of the control strategy employed by the agent, which reduces turbulent kinetic energy near the surface at the rear of the cylinder, thus increasing static pressure at the back, reducing pressure drag, and achieving drag reduction. However, given the limited output power of the jets, a further improvement is needed to modify the flow in the reattachment zone. From the perspective of the dynamic development of shedding vortices, it is clear that when vortices of opposite rotation directions meet, the turbulent kinetic energy in the region decreases, the energy is dissipated, and the vorticity is reduced. 
For future research, it is essential to delve deeper into the flow mechanisms, particularly under high Reynolds number conditions, to elucidate the interactions between jets and vortices of various scales. To this end, transitioning to a three-dimensional model for the geometry is necessary.

\section{Conclusions}\label{sec:conclusion}

The study used the Proximal Policy Optimization (PPO) method from Deep Reinforcement Learning (DRL) as a flow control model and simulations were performed using the $k-\omega$ SST turbulence model. The trained agent successfully reduced drag for a bluff body in high Reynolds number turbulence by periodically controlling the flow field. The results consist of the following. 
\begin{itemize}
    \item This study introduces Deep Reinforcement Learning (DRL) and establishes a zero net flow active flow control (AFC) model to reduce drag for bluff bodies under complex flow conditions at high Reynolds numbers.
    \item Training with random initialization of the flow field for each episode resulted in three distinct training sessions producing stabilized curves within a certain range, demonstrating that the training is reproducible.
    \item Under high Reynolds number conditions, the flow field is very intricate and displays significant nonlinearity. The agent gathered restricted data from the flow field due to potential interference between upstream and downstream probes; therefore, only data regarding velocity and pressure near the cylinder walls was obtained. The trained agent can achieve a fast response and control with a certain degree of robustness. The agent control method successfully decreased the drag on the cylinder by around $30\%$ and showed reliable robustness within a specific range of Reynolds numbers. 
    \item The agent, trained using Active Flow Control (AFC), mainly influenced the high-frequency small-scale structures of the flow field under the current conditions, with little effect on the frequency and energy of vortex shedding. It successfully dampened the low-frequency longitudinal oscillations affecting the cylinder, leading to a decrease in the fluctuating lift forces. 
\end{itemize}




Based on this research, the primary goal for the future is to further explore the use of temporal algorithms such as LSTM (Long-Short Term Memory) to enable the agent to better and more quickly learn the periodic characteristics of the flow, thereby reducing training time costs. On this basis, our aim is to train the agent using simulation data within a \(\pm 10\%-15\%\) range of Reynolds numbers to further enhance the control robustness at high Reynolds numbers. Considering the practical conditions in the experiments, further studies are needed on the limitations with regard to the number of probes and their placement positions.

\begin{acknowledgments}
The present research is part of the activities of the "Dipartimento di Eccellenza 2023-2027".
The second and third authors are members of the "Gruppi Nazionali di Ricerca INdAM".
The authors express sincere thanks to Shumin Li for her helpful insights and contributions.
\end{acknowledgments}

\section*{Data Availability Statement}

Data supporting the findings of this study are available from the corresponding author on a reasonable request.

\appendix
\section{Deep Reinforcement Learning - Proximal Policy Optimization}\label{sec:appendixA}

A Markov Decision Process (MDP) is built based on a set of interactive objects, as shown in Fig.~\ref{fig:3}, called an agent and environment\cite{Sutton2018}. Its elements include states, actions, and rewards. The agent will perceive the current state of the system, implement actions on the environment according to the policy, thus changing the state of the environment, and receiving a reward. In general, the time steps of the MDP can be finite or infinite; for the finite MDP there exists a terminal state which can be defined by episode; in this circumstance, the space of state and action are finite, so is the return. If there is no terminal state, in correspondence with the state, the action and return are infinite. Based on this process, the trajectory is given:

\begin{figure}[h]
  \centering
  \includegraphics[width=0.8\linewidth]{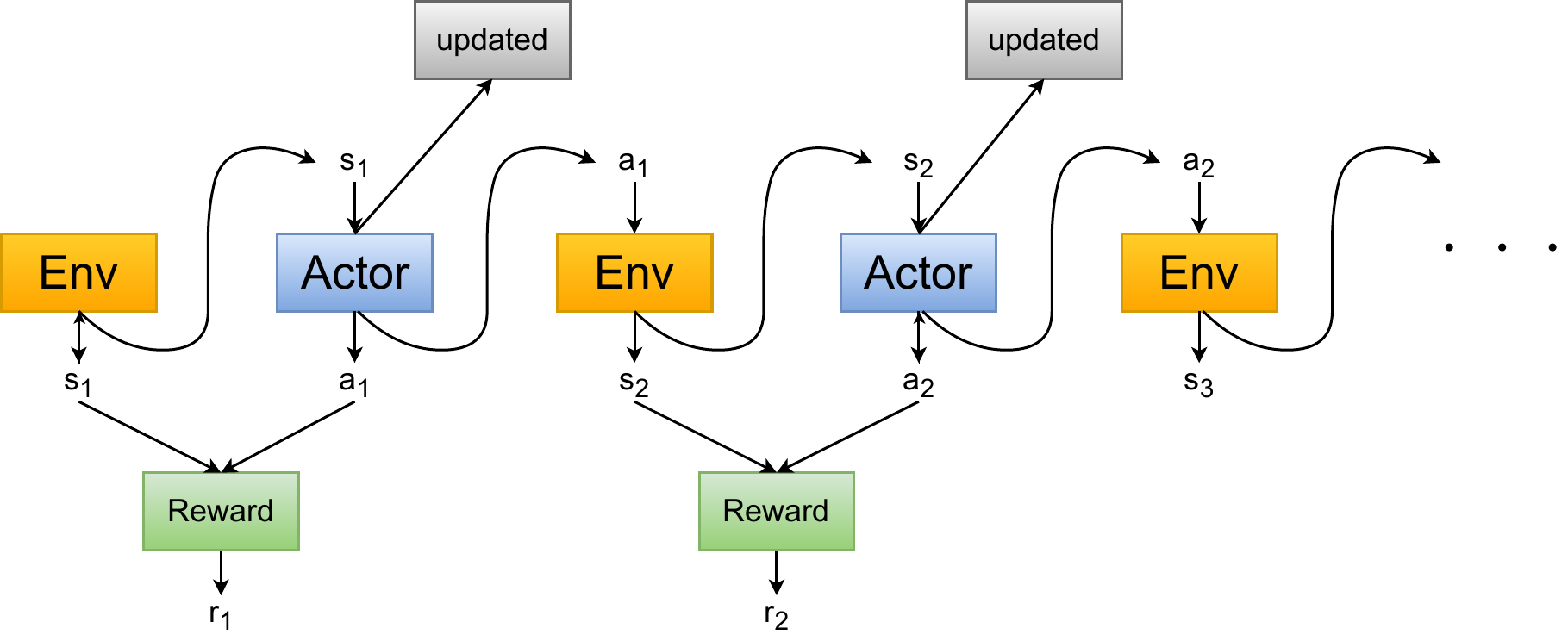}
  \caption{Markov decision process. The agent starts at $S_t$, the actor takes action $A_t$ according to the state $S_t$, then, the environment generates a new state $S_{t+1}$.}
  \label{fig:3}
\end{figure}

\begin{equation}
S_0, A_0, R_0, S_1, A_1, R_1, S_2, A_2, R_2 \dots S_t, A_t, R_t.
\label{eq:9}
\end{equation}
This trajectory describes the problem in a discrete time sequence, so the subscript $t$ means the discrete time or time step.

We introduce the return $G_t$, which is the sum of the rewards $R$ of each discrete time $t$ for an episode;
\begin{equation}
G_t=R_{t+1}+R_{t+2}+\cdot\cdot\cdot+R_T.
\label{eq:10}
\end{equation}
When modeling real problems, according to the definition above, unless $G_t$ tends to converge, it is inappropriate to consider infinite reward. To change this situation, we need to introduce the discounted return for episodic tasks,
\begin{equation}
G_t =R_{t+1}+{\gamma R}_{t+2}+{\gamma^2R}_{t+3}+\dots
 = \sum_{k=0}^{\infty}{\gamma^kR_{t+k+1}=R_{t+1}+{\gamma G}_{t+1}}.
\label{eq:11}
\end{equation}
The discount factor $\gamma$ is a hyperparameter, in the range $[0, 1]$ and indicates importance of future rewards at the current time. If $\gamma=0$, the agent is focused only on the current reward without looking at the future. If $\gamma=1$, it recovers to the undiscounted return. So, using the discounted return, it seems to appear an infinite return in an infinite MDP task and makes the solution of MDP feasible.

The policy is specifically the action in each state, defined by the function $\pi$,   
\begin{equation}
\pi(a|s)\ =\ P(A_t=a|S_t=s).  
\label{eq:12}
\end{equation}
Mathematically, the function $\pi$ is the density of probability. We also define $\pi(a|s,\ \theta)$ as the policy network, which aims to approximate the policy $\pi(a|s)$ by using the policy network $\pi(a|s,\ \theta)$. The property of the policy network satisfies:
\begin{equation}
\sum_{a\in A}{\pi(a|s,\ \theta)=1}.  
\label{eq:13}
\end{equation}
The value function, also called state-value function, mathematically is an expectation of the reward for subsequent actions determined by the policies, which is used to evaluate the quality of the state:
\begin{equation}
V_\pi(s)\ =\ E_\pi[G_t\ |\ S_t=s] = E_\pi[\sum_{k=0}^{\infty}\gamma^kR_{t+k+1}\ |\ S_t=s].
\label{eq:14}
\end{equation}
When the agent starts to train, we want the value function to be maximized by the critic network, and we hope to get as much reward as possible in the shortest possible time, and the value function only depends on the current state. We have another kind of value function, the $Q$-function, also called action-value function, and differently it depends on the current state and the current action. It is the expectation of the reward that can be obtained in each subsequent state in a given state $s$, after taking an action $a$,
\begin{equation}
Q_\pi(s,a)\  =\ E_\pi[G_t\ |\ S_t=s,\ A_t=a]
 = E_\pi[\sum_{k=0}^{\infty}\gamma^kR_{t+k+1}\ |\ S_t=s, A_t=a].
\label{eq:15}
\end{equation}
We can relate the state-value function and the action-value function by following:
\begin{equation}
V_\pi(s)\  =\ \sum_a(E_\pi[\sum_{k=0}^{\infty}\gamma^kR_{t+k+1}\ |\ S_t=s, A_t=a])\ \pi(a|s)
 = \sum_a\pi(a|s)Q_\pi(s,a).
\label{eq:16}
\end{equation}
We can also use $v(s)$ and $q(s,a)$ to represent the approximated state-value function and the approximated action-value function.

If we consider a situation where all the rewards are positive, in this case, whatever the action the agent takes, to increase the expectation of the value faction, we need to increase the probability of this token action. Thus, it is extremely necessary to introduce the advantage function:
\begin{equation}
A_\pi(s,a)=\ Q_\pi(s,a)\ -\ V_\pi(s).  
\label{eq:17}
\end{equation}
The advantage function provides a standard for evaluating the policy. If the advantage function is positive, it means that in state $s$, taking a new action $a$ is better than the action given by the current policy, then we need to increase the probability of this new action in the RL algorithm by maximizing the expectation of the advantage function in the critic network.
\begin{figure}[h]
    \centering
    \includegraphics[width=0.9\linewidth]{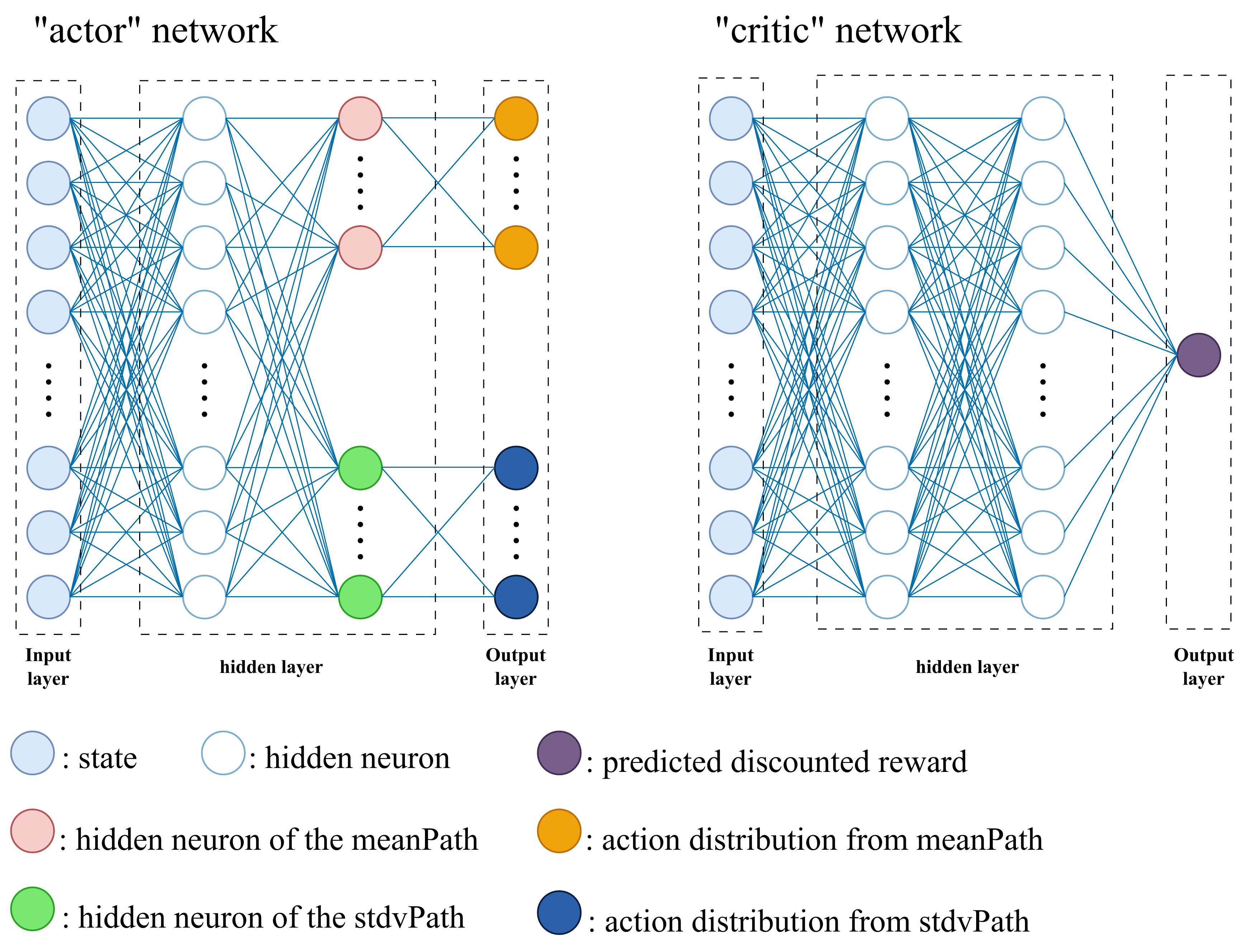}
    \caption{Schematic diagram of actor-critic network used in the PPO algorithm.}
    \label{fig:4}
\end{figure}
The critic network in Fig.~\ref{fig:4} is used to approximate the advantage function of~\cref{eq:17}. In this scenario, the network's input is made up of 768 neurons (each of the 3 layers utilizes 128 probes to measure two parameters-pressure and velocity-for gathering the necessary information). The actor and critic networks both feature four hidden layers, each matching the size of the network's input. The critic network's output includes a single neuron, whereas the actor network's two paths produce outputs corresponding to the number of jets, which is nine for the mean and nine for the standard deviation in \cref{eq:25}. For updating the policy network (actor network), generally we choose a type of gradient descent method to find the parameter $\theta$ for the network, and it is necessary to calculate the gradient of the objective function,
\begin{equation}
\mathrm{\nabla}_\theta J(\theta)\ =\ \mathrm{\nabla}_\theta \{ E_{S\sim\eta, \ A\sim\pi\ }\left[\log{\pi}(A|S,\theta_t)q_\pi(S,A)\right] \} 
{\ =E}_{S\sim\eta,\ A\sim\pi\ }[\mathrm{\nabla}_\theta \log{\pi}(A|S,\theta_t)q_\pi(S,A)].
\label{eq:18}
\end{equation}
Here we introduce a baseline $b(S)$, which is a scalar function of the state space $S$. 
\begin{equation}
\mathrm{\nabla}_\theta J(\theta)\ = E_{S\sim\eta,\ A\sim\pi\ } \{ \mathrm{\nabla}_\theta \log{\pi}(A|S,\theta_t)[q_\pi(S,A)-b(S)] \}.
\label{eq:19}
\end{equation}
Then we need to demonstrate that the introduction of $b(S)$ does not change the gradient value.
\begin{equation}
\begin{aligned}
\mathrm{\nabla}_\theta J(\theta)\ &=E_{S\sim\eta,\ A\sim\pi\ }[\mathrm{\nabla}_\theta \log{\pi}(A|S,\theta_t)b(S)] \\
& = \sum_{s\in S}{\eta(s)} \sum_{a\in A}\pi(a|s,\theta_t)\mathrm{\nabla}_\theta \log{\pi}(a|s,\theta_t)b(s) \\
& = \sum_{s\in S}{\eta(s)} \sum_{a\in A}\mathrm{\nabla}_\theta \pi(a|s,\theta_t)b(s) \\
& = \sum_{s\in S}{\eta(s)}b(s)\sum_{a\in A}{\mathrm{\nabla}_\theta\pi(a|s,\theta_t)} \\
& = \sum_{s\in S}{\eta(s)}b(s)\mathrm{\nabla}_\theta\sum_{a\in A}{\pi(a|s,\theta_t)} \\
& = \sum_{s\in S}{\eta(s)}b(s)\mathrm{\nabla}_\theta1\ =\ 0.
\end{aligned}
\label{eq:20}
\end{equation}
In this case, we choose the value function:
\begin{equation}
  b(S)=\ E_{A\sim\pi}[q(s,A)] = v(s).
\label{eq:21}
\end{equation}
It is difficult or maybe impossible to calculate the expectation as we said before, so we need to resort to a Monte Carlo estimation, but particularly, in this case, as an off-line method, the data set we sample from can be changed.  
\begin{equation}
E_{x\sim p_0}[x] \approx \bar{f} = \frac{1}{n}\sum_{i=1}^nf(x_i)=\frac{1}{n}\sum_{i=1}^n\frac{p_0(x_i)}{p_1(x_i)}x_i.
\label{eq:22}
\end{equation}
Usually, for the PPO algorithm, we sample from the old policy, combined with the advantage function, and we have the policy gradient:
\begin{equation}
\mathrm{\nabla}_\theta J(\theta) = \ E [\frac{\pi_{new}(A|S,\theta)}{\pi_{old}(A|S,\theta)}\mathrm{\nabla}_\theta \log{\pi_{new}}(A|S,\theta) (q_\pi(S,A)-v_\pi(S))].
\label{eq:23} 
\end{equation}
The update of the parameter $\theta$ is:
\begin{equation}
\theta_{t+1}=\ \theta_t+\ \alpha_\theta\frac{\pi_{new}(a_t|s_t,\theta_t)}{\pi_{old}(a_t|s_t,\theta_t)}\mathrm{\nabla}_\theta \log{\pi_{new}}(a_t|s_t,\theta_t)
(q_t(s_t,a_t)-v_t(s_t)).
\label{eq:24} 
\end{equation}
After updating the parameter, we can get the action distribution at the current observation (states) from the actor network. The distribution of the action can be written as: 
\begin{equation}
p\left(a\right)=\frac{1}{\xi\sqrt{2\pi}}e^{-\ \frac{{(\alpha-\mu)}^2}{2\xi^2}},
\label{eq:25} 
\end{equation}
where $\xi$ is the standard deviation, and $\mu$ is the mean value; these two parameters are the outputs of the actor network. Having this distribution, the agent chooses an action as an input into the environment.

For avoiding a destructive large weight update, we introduce the PPO clip function:
\begin{equation}
L^{CLIP}(\theta)= E[\min \{ r(\theta)\hat{a}_\pi, \textrm{clip}(r(\theta ), 1-\epsilon, 1+\epsilon) \hat{a}_\pi\}],
\label{eq:26} 
\end{equation}
\begin{eqnarray}
&{\hat{a}}_\pi = q_\pi(S,A)-v_\pi(S), \\ \label{eq:27}
&r\left(\theta\right) = \displaystyle \frac{\pi_{new}(a_t|s_t,\theta_t)}{\pi_{old}(a_t|s_t,\theta_t)},
\label{eq:28}
\end{eqnarray}
where ${\hat{a}}_\pi$ is the approximate advantage function. There are two possible cases depending on the positivity of the advantage function. The first is when the advantage function is positive, when $L^{CLIP}$ turns to:
\begin{equation}
    L^{CLIP}\left(\theta\right)=E[\min \{r\left(\theta\right),\ \left(1+\epsilon\right)\}{\hat{a}}_\pi].
    \label{eq:29}
\end{equation}
In this case, the object is to maximize $L^{CLIP}\left(\theta\right)$, giving the approximate advantage function, thus we need to maximize the ratio $r\left(\theta\right)$ between the new policy and the old policy, but if the difference is large, it makes no sense to the physical environment. So, with this limiting function, the ratio will reach a ceiling $(1+\epsilon){\hat{a}}_\pi$.
The second case is when the advantage function is negative, where the PPO clip function becomes:
\begin{equation}
L^{CLIP}\left(\theta\right)=E[\max \{r\left(\theta\right),\ \left(1-\epsilon\right)\}{\hat{a}}_\pi].
\label{eq:30}
\end{equation}
\section{Hyperparameters PPO}
\label{sec:appendixB}
The following are the hyperparameters used for training the PPO agent in this study:
\begin{table}[h]
    \centering
    \begin{tabular}{|c|c|}
        \hline
        Minibatch size & 64 \\ \hline
        Experience horizon & 64\\ \hline
        Discount factor & 0.97\\ \hline
        Clip factor & 0.5\\ \hline
        Entropy loss weight & 0.03\\ \hline
        Number of epochs & 3\\ \hline
        Learning rate & 0.001 \\ \hline
      \end{tabular}
    \caption{PPO hyperparameters.}
    \label{tab:Hyperparameters PPO}
\end{table}


\bibliography{aipsamp}

\end{document}